\documentclass[reprint,aps,pra,superscriptaddress,nofootinbib,floatfix]{revtex4-2}

\usepackage{amsmath,amssymb}
\usepackage{graphicx}
\usepackage{xcolor}
\usepackage{hyperref}
\hypersetup{colorlinks=true,linkcolor=blue!60!black,citecolor=blue!60!black,urlcolor=blue!60!black}

\newcommand{\PhiP}{\Phi^{+}}

\newtheorem{proposition}{Proposition}
\newtheorem{theorem}{Theorem}
\newtheorem{corollary}{Corollary}
\newtheorem{lemma}{Lemma}

\begin{document}

\title{Operational thresholds of Bell mixtures with complex X noise}

\author{Xuan Du Trinh}
\email{xtrinh@cs.stonybrook.edu}
\affiliation{Stony Brook University, Stony Brook, New York 11794, USA}
\date{August 18, 2026}

\begin{abstract}
Noisy Bell mixtures may carry entanglement, standard teleportation
usefulness, projective-measurement steerability, optimized
Cavalcanti--Jones--Wiseman--Reid (CJWR) violation, and optimized
Clauser--Horne--Shimony--Holt (CHSH) violation. If the noise state already possesses one of these abilities,
the mixture may lose and later recover it at distinct boundary crossings as the
Bell weight decreases from $1$ to $0$. We characterize the corresponding
ability-absence intervals and definitive thresholds for mixtures of
$|\Phi^+\rangle$ with an arbitrary complex two-qubit $X$ noise state
$\sigma_X$. A closed-form
singular-value flow of the Pauli correlation tensor yields the exact teleportation-useless,
CJWR-satisfying, and CHSH-local intervals, while the positive partial
transpose (PPT) criterion yields the exact separable interval. At fixed populations and
coherence magnitudes of $\sigma_X$, these intervals widen monotonically as the
relative phase between the $\Phi$-block coherence of $\sigma_X$ and the Bell
coherence increases from $0$ (aligned) to $\pi$ (anti-aligned). The definitive
thresholds obey a universal ordered chain
from the entanglement threshold through the teleportation usefulness threshold
and the three-setting CJWR witness threshold to the common two-setting CJWR
witness and CHSH-nonlocal threshold. A second chain is obtained by replacing
the teleportation usefulness threshold with the steerability threshold. The
teleportation usefulness and steerability thresholds are not mutually ordered.
A state may be useful for standard teleportation but unsteerable, or
steerable but not useful for standard teleportation. For arbitrary pure
two-qubit noise, the steerability
thresholds in both directions for projective measurements and arbitrary
positive operator-valued measures (POVMs) all equal the entanglement threshold.
The same equality holds for two diagonal mixed $X$ families, for which the
common threshold is zero. For arbitrary product noise, an effective $X$-state
singular-value flow determined by the product of the local Bloch radii and one
relative Bell-frame angle gives the teleportation usefulness, CJWR witness, and
CHSH-nonlocal thresholds. If either local noise state is pure, the entanglement
threshold and all four steerability thresholds are zero. For generic full-rank
mixed $X$ noise and full-rank product noise, exact directional steerability thresholds
remain unknown, and finite-setting semidefinite programs give upper bounds on
the projective-measurement steerability thresholds.
\end{abstract}

\maketitle

\predisplaypenalty=100\relax

\section{Introduction}
\label{sec:intro}

Maximally entangled qubit pairs are basic resources for quantum
communication, but imperfections in preparation, transmission,
storage, and local control replace ideal Bell pairs with noisy states.
For such a pair, the relevant question is not only whether it remains
entangled, but which tasks it can still support. Can it teleport a
qubit better than any classical strategy~\cite{Popescu1994,
HorodeckiTeleportationBell1996}? Can one party demonstrate steering of
the other's state with the available measurements? Can its correlations
violate the CHSH bound after optimization over local measurement
settings? These abilities are distinct: a state may be entangled but
teleportation-useless, teleportation-useful but not certified by a
chosen steering witness, or steerable while remaining
CHSH-local~\cite{JonesWisemanDoherty2007,BrunnerEtAl2014,
UolaCostaNguyenGuhne2020}.
As the noise admixture changes, these operational abilities may appear
or disappear at different stages, simultaneously, not at all, or in a
loss-and-recovery sequence.
They also distinguish trust regimes in quantum key distribution.
Entanglement-based protocols use characterized devices, steering can
support one-sided device-independent security, and CHSH violation can
support fully device-independent security~\cite{CurtyLewensteinLutkenhaus2004,
BranciardEtAl2012,AcinEtAl2007}.

To study these changes at different noise levels, we consider the line
\begin{equation}
\rho_\lambda=\lambda\PhiP+(1-\lambda)\sigma_X,
\qquad 0\le\lambda\le1,
\label{eq:mixing-line}
\end{equation}
where $\PhiP=|\PhiP\rangle\!\langle\PhiP|$ is the reference Bell-state
density operator defined by
$|\PhiP\rangle=(|00\rangle+|11\rangle)/\sqrt2$. The parameter $\lambda$ is
the Bell weight, and $\sigma_X$ is an arbitrary physical two-qubit $X$ state
whose coherences may be complex. The line can be read in two
complementary directions. If one reads the line from $\lambda=0$, increasing
$\lambda$ adds Bell-state weight and tracks the operational abilities present
in $\rho_\lambda$. On the other hand, if one reads the line from $\lambda=1$,
moving toward $\lambda=0$ adds noise weight and tracks the loss and possible
recovery of operational abilities.
The problem addressed here is not the evaluation of one resource in
isolation, but a unified description of how the operational
thresholds are ordered and crossed along this line.

In the computational basis, the $X$ noise state is
\begin{equation}
\sigma_X=
\begin{pmatrix}
a&0&0&u\\
0&b&v&0\\
0&v^*&c&0\\
u^*&0&0&d
\end{pmatrix},
\qquad
\bigl(a,b,c,d\ge0,\;\;u,v\in\mathbb C\bigr).
\label{eq:x-matrix}
\end{equation}

The $X$ family includes physically important and analytically
tractable classes, including Werner and Bell-diagonal states. It is also
preserved under common noise mechanisms and network operations, including
local pure dephasing, local amplitude damping, effective Pauli channels, and
post-selected entanglement swapping based on Bell-state
measurements~\cite{ZukowskiEtAl1993,MunozGruningRoa2014}
(Sec.~\ref{sec:death-times}). The $X$ family is widely used in models of
two-qubit open-system dynamics and entanglement sudden
death~\cite{YuEberly2007}, and often permits closed expressions for two-qubit
correlations~\cite{AliRauAlber2010,QuesadaAlQasimiJames2012,Hu2013XStates}.
For either
concurrence~\cite{Wootters1998} or
negativity~\cite{VidalWerner2002}, every two-qubit state also has an $X$-state counterpart
with the same spectrum and the same value of that entanglement
measure~\cite{MendoncaEtAl2014}. Earlier studies
showed that the initial $X$-state coherences can change entanglement
dynamics and the onset of sudden death~\cite{NunavathMishraPathak2022}.
Motivated by these results, we investigate how the populations and
coherences of $\sigma_X$ control the threshold crossings for
entanglement, teleportation usefulness, steering, and Bell nonlocality
along the mixing line of Eq.~\eqref{eq:mixing-line}.
We call the corresponding thresholds the entanglement threshold,
teleportation usefulness threshold, steerability threshold, and
CHSH-nonlocal threshold, respectively. Unless another measurement class
is specified, steerability refers to projective-measurement steerability.

For Bell nonlocality, we use the CHSH
criterion~\cite{CHSH1969,HorodeckiCHSH1995}. For steering, we distinguish
finite-setting Cavalcanti--Jones--Wiseman--Reid (CJWR)
witnesses~\cite{CavalcantiJonesWisemanReid2009} from the exact steerability
boundary for the full class of projective measurements.
Projective steering is directional and
excludes a local-hidden-state model for the conditional states of the
trusted party~\cite{WisemanJonesDoherty2007,BrunnerEtAl2014,
UolaCostaNguyenGuhne2020}. A CJWR violation certifies projective
steering, but nonviolation does not determine whether the state is steerable
using projective measurements. We
call a state violating no $n$-setting CJWR inequality
\emph{CJWR-$n$-satisfying}.

The different thresholds require complementary calculations. The
positive partial transpose (PPT) criterion locates the separability
boundary, and the fully entangled fraction determines standard
teleportation usefulness.
The optimized CHSH and CJWR values depend on
the singular values of the Pauli correlation tensor
$T_{ij}(\lambda)=\langle\sigma_i\otimes\sigma_j\rangle_{\rho_\lambda}$,
$i,j\in\{x,y,z\}$~\cite{HorodeckiCHSH1995,CostaAngelo2016}.

The main mathematical tool for the witness thresholds is the closed-form
singular-value flow of $T(\lambda)$. For real $X$ noise, $T(\lambda)$ is
diagonal, and its three entries are affine in $\lambda$. For complex
coherences, the $xy$ cross correlators need not vanish. A local basis
change by the opposite phase rotations
$W_\phi=\mathrm{diag}(1,e^{i\phi})\otimes
\mathrm{diag}(1,e^{-i\phi})$ leaves $\PhiP$ invariant and maps
$v\mapsto e^{-2i\phi}v$. Choosing $\phi=\arg v/2$ makes $v$ real and
nonnegative. In this gauge, $\theta:=\arg u$ is the phase of the noise
coherence $u$ relative to the real, positive Bell coherence. We derive
the three singular values in closed form:
\begin{equation*}
s_\pm(\lambda)=|Q(\lambda)\pm P(\lambda)|,
\qquad |t_z(\lambda)|,
\end{equation*}
where
\begin{align*}
Q(\lambda)&=|\lambda+2(1-\lambda)u|,
&
P(\lambda)&=2(1-\lambda)v,\\
t_z(\lambda)&=\gamma+\lambda(1-\gamma),
&
\gamma&=a-b-c+d.
\end{align*}
Thus, all dependence of the tensor thresholds on the relative phase
$\theta$ enters through $Q$ (Proposition~\ref{prop:flow}).

The operational abilities considered here are entanglement, standard
teleportation usefulness, projective-measurement steerability, optimized three-setting
CJWR violation, and optimized CHSH violation.
Let $\mathcal O$ denote one of these abilities. The set of
states that lack $\mathcal O$ is convex. For example, the separable states
form the convex set of states that lack entanglement.
Since $\rho_\lambda$ is affine in $\lambda$, convexity implies
that the Bell weights for which $\rho_\lambda$ lacks $\mathcal O$ form a single
interval.
The interval may contain $\lambda=0$, be detached from that endpoint,
consist of a single point, or be empty. We define the
\emph{definitive threshold} as the upper edge of this interval when it is
nonempty and as zero when it is empty (Sec.~\ref{sec:xsetup}).
An interval containing $\lambda=0$ describes a single onset of
$\mathcal O$ at its upper edge. When the noise state already has $\mathcal O$
and the corresponding interval is nonempty, that interval is detached from
$\lambda=0$. For a detached interval with two distinct edges,
$\rho_\lambda$ has $\mathcal O$ below the lower edge, lacks $\mathcal O$
throughout the interval, and has $\mathcal O$ again above the upper edge.

At fixed populations and coherence magnitudes, the relative phase
$\theta$ is the only parameter that the extension to the full complex
$X$-noise family effectively adds to the real $X$-noise family. Studying its
effect on the ability-absence intervals and the definitive thresholds therefore
shows directly what changes under this extension.

{The main results are organized as follows.}

\begin{figure*}[!t]
\centering
\includegraphics[width=0.98\textwidth]{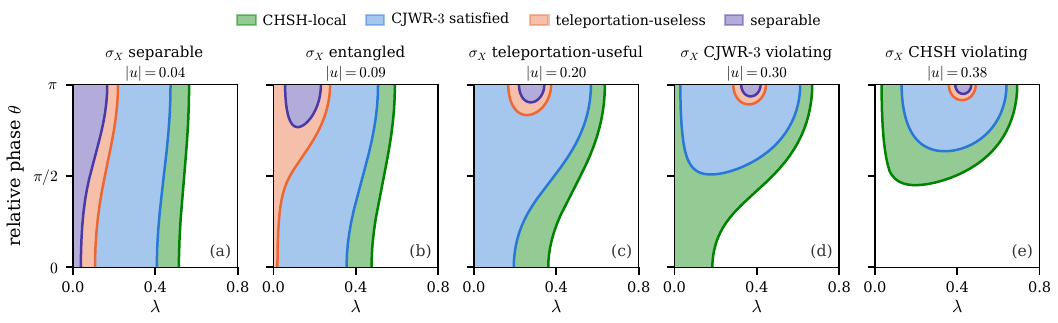}
\caption{Operational phase diagrams of the Bell-mixing line
$\rho_\lambda$ in the $(\lambda,\theta)$ plane. All panels use
$(a,b,c,d)=(0.42,0.18,0.02,0.38)$ and $v=0.04$, with the value of $|u|$
shown above each panel. As $|u|$ increases, $\sigma_X$ is (a)~separable,
(b)~entangled but teleportation-useless, (c)~teleportation-useful but
CJWR-$3$-satisfying, (d)~CJWR-$3$-violating but CHSH-local, and
(e)~CHSH-violating. Colors mark where $\rho_\lambda$ lacks each ability:
green for CHSH locality, blue for CJWR-$3$ satisfaction, orange for
teleportation uselessness, and violet for separability. These regions nest
in the same order. A region touching $\lambda=0$ indicates a single onset for
an ability absent from $\sigma_X$. A detached region indicates that an ability
present at $\lambda=0$ is lost at its lower edge and recovered at its
definitive upper edge. Across panels, the regions detach according to the
threshold hierarchy. In every panel, each region widens
monotonically as $\theta$ increases from $0$ to $\pi$
(Appendix~\ref{app:mono}).}
\label{fig:window-widening}
\end{figure*}

\paragraph*{Analytic results.}
The PPT criterion gives the exact separable interval. From the
singular-value flow, we obtain the fully entangled fraction and hence the exact
teleportation-useless interval, together with the exact CJWR-$3$-satisfying and
CHSH-local intervals. The CHSH-local interval is also the
CJWR-$2$-satisfying interval. The CJWR-$3$ boundary points are roots of an
explicit quadratic equation (Sec.~\ref{sec:cjwr-main}). The CHSH boundary
points are roots of one quadratic equation and one quartic equation
(Sec.~\ref{sec:chsh-main}, Appendix~\ref{app:chsh}).

For separable $\sigma_X$, Ref.~\cite{eac-paper} derived the single-onset
entanglement and teleportation usefulness thresholds from the PPT criterion
and the fully entangled fraction. For entangled $\sigma_X$, the same work
proved that the separable and teleportation-useless subsets of the Bell-mixing
line each form a single interval when nonempty, but it did not derive their
endpoints. Our boundary equations give the complete intervals for every
physical $X$ noise state and recover the single-onset formulas when $\sigma_X$
is separable.

At fixed populations and coherence magnitudes, the boundary conditions above
also establish phase widening of the separable, teleportation-useless,
CJWR-$3$-satisfying, and CHSH-local intervals. As $\theta$ increases from $0$
to $\pi$, their lower edges cannot rise and their upper edges cannot fall, so
the intervals widen monotonically. Their upper edges are the definitive
thresholds and are therefore nondecreasing in $\theta$. Hence, the two
real-coherence endpoints bracket the thresholds.
Figure~\ref{fig:window-widening} illustrates the nesting, detachment, and phase
widening of these intervals for five example $X$ noise states. For
ability-absence intervals whose boundaries are not known in closed form, phase
widening is studied in Ref.~\cite{PhaseOrderingPRL}, which also
treats threshold bounds for noise beyond the $X$ family.

\paragraph*{Structural results.}
We determine how the definitive thresholds of the operational abilities are
ordered across all noise states. For each pair of abilities, we establish
whether one threshold is dominated by the other for every noise state or
whether their order depends on the noise state
(Corollary~\ref{cor:order}). In particular, neither standard teleportation
usefulness nor projective-measurement steerability implies the other. On the
Werner line, where $\sigma_X=I_4/4$, $\rho_\lambda$ is useful for standard
teleportation but unsteerable under projective measurements when
$1/3<\lambda\le1/2$~\cite{Werner1989,JonesWisemanDoherty2007}. Conversely,
when $\sigma_X=|01\rangle\langle01|$, $\rho_\lambda$ is steerable in both
directions using projective measurements but not useful for standard
teleportation when $0<\lambda\le1/2$. Thus, depending on the noise state, the
teleportation usefulness threshold may be smaller or larger than the
steerability threshold.

\paragraph*{Exact steerability results.}
For a two-qubit noise state $\sigma$, write the Bell mixture as
$$
\rho_\lambda
=\lambda\PhiP
+(1-\lambda)\sigma.
$$
When $\sigma=|\psi\rangle\langle\psi|$ is any pure two-qubit
state, {not necessarily of $X$ form,} we prove that, for every
$\lambda\in[0,1]$, $\rho_\lambda$ is
entangled if and only if it is steerable in both directions using projective
measurements~{(Theorem~\ref{cor:pure-noise-steering})}.
{For mixed $X$ noise, we consider the two diagonal families
$\sigma_X=\operatorname{diag}(a,b,0,d)$ and
$\sigma_X=\operatorname{diag}(a,0,c,d)$. Every mixture in either family with
positive Bell weight is entangled and steerable in both directions using
projective measurements{, so the entanglement and the steerability
thresholds are all equal to zero
(Proposition~\ref{prop:boundary-steering})}. Across these three families of noise states,
allowing arbitrary positive operator-valued measures (POVMs) offers no
advantage over projective measurements for steering in both directions along
the Bell mixture~{(Corollary~\ref{cor:povm-projective})}.}
Outside the exact families described above, no general formula for the
projective-measurement steerability threshold is known. For generic full-rank
mixed $X$ noise, we therefore use semidefinite programs to obtain
finite-setting bounds. We run the programs with eight and sixteen measurement
directions and compare the resulting bounds with the optimized CJWR-$3$
witness threshold and the CJWR-$2$ witness threshold, which equals the
CHSH-nonlocal threshold (Sec.~\ref{sec:true-steering}). Special exact
full-rank examples nevertheless exist. Ref.~\cite{SteeringMarginalsLetter}
gives product-noise and entangled-$X$-noise examples with the same correlation
tensor but different $A\to B$ projective-measurement steerability thresholds.

\paragraph*{Product-noise reduction.}
{Although a general product noise state need not have $X$ form,
the singular-value flow of the corresponding Bell mixture can be obtained
from an effective complex-$X$ line.}
For arbitrary product noise $\sigma=\sigma_A\otimes\sigma_B$, we reduce the
CHSH-nonlocal threshold and CJWR witness thresholds to two parameters: the
product $g=r_Ar_B$ of the local Bloch-vector lengths and one relative Bell-frame angle $\varphi$
(Sec.~\ref{sec:product-law}). This reduction complements
{Ref.~\cite{eac-paper}, which derives the entanglement and
teleportation usefulness thresholds for arbitrary product noise and for
separable $X$ noise.} {Together, the two analyses characterize the
entanglement, teleportation usefulness, CJWR-$3$ witness, and CHSH-nonlocal
thresholds for arbitrary product noise. We obtain closed-form expressions
for the first three, while explicit quadratic and quartic boundary equations
determine the CHSH-nonlocal threshold, which is also the CJWR-$2$ witness
threshold.}
{When at least one of $\sigma_A$ and $\sigma_B$ is pure, the same section
proves that, for every $\lambda\in[0,1]$, $\rho_\lambda$ is entangled if and
only if it is steerable in both directions using projective measurements.
POVMs again offer no advantage over projective measurements for steering in
both directions along the Bell mixture
(Proposition~\ref{prop:product-pure-factor}). For full-rank product noise,
however, the correlation-tensor reduction does not determine the directional
projective-measurement steerability thresholds~\cite{SteeringMarginalsLetter}.}

\paragraph*{Dynamical applications.}
We also apply the closed-form formulas along two trajectories that keep
the state within the $X$ family: time evolution under local noise channels and a
post-selected entanglement-swapping
chain~\cite{ZukowskiEtAl1993,MunozGruningRoa2014}. Along these trajectories,
the formulas track when each operational
ability is lost (Sec.~\ref{sec:death-times}, Fig.~\ref{fig:death}).

{The rest of the paper is organized as follows.}
Section~\ref{sec:xsetup} motivates the mixing line for an arbitrary complex
two-qubit $X$ noise state $\sigma_X$ and derives its closed-form singular-value
flow. It establishes the ability-absence intervals and defines the definitive
thresholds.
Sections~\ref{sec:chsh-main} and~\ref{sec:cjwr-main} derive the
CHSH-nonlocal threshold and CJWR witness thresholds.
{Section~\ref{sec:true-steering} proves the exact steering results.
Section~\ref{sec:alignment} establishes relative-phase monotonicity of
the definitive thresholds at fixed populations and coherence magnitudes.
Section~\ref{sec:product-law} treats the thresholds for product noise.}
Section~\ref{sec:death-times} presents the
dynamical applications, and Sec.~\ref{sec:conclusion}
summarizes the results in Table~\ref{tab:thresholds}, discusses the
limitations of correlation-tensor information, and concludes.

\section{Bell-state mixtures of complex X states}
\label{sec:xsetup}

This section motivates the mixing line of Eq.~\eqref{eq:mixing-line} between
$\PhiP$ and an arbitrary complex two-qubit $X$ noise state
$\sigma_X$. The line runs from the noise state at $\lambda=0$
to the Bell state at $\lambda=1$ and can be read in two complementary
directions. If one reads the line from $\lambda=0$, increasing $\lambda$ adds
Bell-state weight and tracks the operational abilities present in
$\rho_\lambda$. On the other hand, if one reads the line from $\lambda=1$,
moving toward $\lambda=0$ adds noise weight and tracks the loss and possible
recovery of operational abilities. After
stating the physical constraints on the complex two-qubit $X$ family and
showing how the line arises from stochastic noise, we introduce a gauge that
allows us to remove one coherence phase through a local rotation while leaving
the physics unchanged. In this gauge, we derive the closed-form singular-value
flow (Proposition~\ref{prop:flow}). Finally,
Proposition~\ref{prop:windows} establishes that the weights for which
$\rho_\lambda$ lacks a given operational ability form a single interval. When
the interval is nonempty, its upper edge defines the definitive threshold,
while an empty interval gives zero by convention.

\paragraph*{$X$ family and our hypotheses.}
The density-matrix conditions on the entries of
Eq.~\eqref{eq:x-matrix} are
\begin{equation}
a+b+c+d=1,\qquad |u|^2\le ad,\qquad |v|^2\le bc,
\label{eq:x-density-conditions}
\end{equation}
together with $a,b,c,d\ge0$.
The $X$ noise state $\sigma_X$ is separable if and only if it satisfies
the positive partial transpose (PPT) criterion~\cite{Peres1996,Horodecki1996}:
\begin{equation}
|u|^2\le bc,\qquad |v|^2\le ad.
\label{eq:x-separability-conditions}
\end{equation}
We do not assume separable noise. The only statements below that
require separable noise are the readings of the entanglement and
teleportation usefulness thresholds as single onsets, fixed by the
absorption capacities of Ref.~\cite{eac-paper}. Each such statement
says so explicitly. For entangled noise, the same
PPT criterion determines all separability--entanglement boundary crossings.
The fully entangled fraction $f$, the maximal overlap of the state
with a maximally entangled state~\cite{HorodeckiTeleportation1999}, determines
all teleportation usefulness boundary crossings.

\paragraph*{Physical origin of the line.}
Both the mixing line and the $X$ form arise naturally in a stochastic
noise model. The Bell state $\PhiP$ is itself an $X$ state.
Local pure dephasing and local amplitude damping preserve the $X$ family.
More generally, any local channel whose Kraus operators are each diagonal or
antidiagonal in the computational basis preserves the $X$ family
(Appendix~\ref{app:channels}). This class includes every qubit Pauli channel.
Suppose a pair is left undisturbed with probability
$\lambda$ and otherwise undergoes an $X$-preserving channel
$\mathcal E$. Let
$\sigma_X=\mathcal E(\PhiP)$ denote the noisy output.
The effective channel $\lambda\,\mathrm{id}+(1-\lambda)\mathcal E$ then maps
$\PhiP$ to the state $\rho_\lambda$ defined in
Eq.~\eqref{eq:mixing-line}. If the channel strength or exposure time varies
between experimental runs, the ensemble-averaged output remains an $X$ state
because the $X$ family is convex.

The state $\sigma_X$ can be entangled or separable. Starting
from the entangled state $\PhiP$, neither pure dephasing nor
amplitude damping alone reaches a separable output at any finite exposure.
Pure dephasing shrinks the coherence $u$ and changes nothing else, so the
state stays entangled until the coherence is fully dephased. Amplitude
damping keeps the state entangled until the limit of complete amplitude
damping.
Only their combination can make the state separable after a finite exposure
time.

Concretely, each qubit relaxes with amplitude-damping time $T_1$, the
mean lifetime of $|1\rangle$, and dephases with pure-dephasing time
$T_\varphi$. After an exposure time $t$, a qubit in $|1\rangle$ remains there with
probability $\gamma_1=e^{-t/T_1}$. Pure dephasing, driven by phase fluctuations
with no energy exchange, retains a fraction $\gamma_\varphi=e^{-2t/T_\varphi}$ of
the coherence between $|00\rangle$ and $|11\rangle$ and leaves the
populations fixed~\cite{Krantz2019}.\footnote{On each qubit, amplitude
damping is the channel
$\mathcal A_{\gamma_1}(\rho)=\sum_{j=0}^1 A_j\rho A_j^\dagger$, where
$A_0=\mathrm{diag}(1,\sqrt{\gamma_1})$ and
$A_1=\sqrt{1-\gamma_1}\,|0\rangle\langle1|$. Pure dephasing is the channel
$\mathcal D_{\gamma_\varphi}(\rho)=(1-p)\rho+p\sigma_z\rho\sigma_z$, where
$p=(1-\sqrt{\gamma_\varphi})/2$~\cite{NielsenChuang2010}. The latter scales
each qubit's coherence by $\sqrt{\gamma_\varphi}$ and the pair coherence by
$\gamma_\varphi$. Applying both channels to each qubit of
$\PhiP$ yields $\sigma_X$. Their order is immaterial because
conjugation by $\sigma_z$ sends each amplitude-damping Kraus operator to
$\pm$ itself (Appendix~\ref{app:channels}).} The noisy output is the $X$ state
$$
\sigma_X=\frac12
\begin{pmatrix}
1+(1-\gamma_1)^2 & 0 & 0 & \gamma_\varphi\gamma_1\\
0 & (1-\gamma_1)\gamma_1 & 0 & 0\\
0 & 0 & (1-\gamma_1)\gamma_1 & 0\\
\gamma_\varphi\gamma_1 & 0 & 0 & \gamma_1^2
\end{pmatrix},
$$
whose concurrence~\cite{Wootters1998} is
$C(\sigma_X)=\max\{0,\gamma_1(\gamma_\varphi+\gamma_1-1)\}$, so the state is entangled only
while $\gamma_\varphi+\gamma_1>1$ and its entanglement falls to zero at a finite
exposure, the sudden-death mechanism discovered by Yu and
Eberly~\cite{YuEberly2004}. Within this model, short exposures produce
entangled $X$ noise, whereas sufficiently long exposures produce
separable $X$ noise. Thus both cases arise from the same physical
process.

More generally, $\sigma_X$ is an arbitrary physical two-qubit $X$ state with
possibly complex coherences. The four Bell states split into two blocks:
the $\Phi$ block $\{|00\rangle,|11\rangle\}$, whose coherence is $u$,
and the $\Psi$ block $\{|01\rangle,|10\rangle\}$, whose coherence is
$v$. A pure-dephasing or amplitude-damping channel acting on
$\PhiP$ leaves $u$ real and nonnegative. A source that
instead emits a $\Psi$-block state, such as a singlet or a
post-selected Bell-measurement swap heralding $|\Psi^\pm\rangle$,
produces the $\Psi$-block coherence $v$.
Uncompensated single-qubit phase rotations $\mathrm{diag}(1,e^{i\phi})$,
arising, for example, from a residual detuning or a miscalibrated
phase gate,
multiply $u$ and $v$ by phase factors and can make these coherences
complex.

\paragraph*{Gauge reduction.}
The next question is which parameters of $\sigma_X$ the thresholds
actually depend on. The local phase rotation
$W_\phi=\mathrm{diag}(1,e^{i\phi})\otimes
\mathrm{diag}(1,e^{-i\phi})$ applies opposite phase rotations to the
two qubits. It leaves $\PhiP$ and $u$ unchanged and maps
$v\mapsto e^{-2i\phi}v$ (Appendix~\ref{app:gauge}). Because $W_\phi$ is a local
unitary, it changes only the local bases and cannot change whether
$\rho_\lambda$ possesses any of the bipartite operational abilities considered
here. Applying $W_\phi$ along the entire line therefore leaves every
operational threshold unchanged. Choosing
$\phi=\arg v/2$ makes $v\ge0$ without loss of generality.
The phase
$$
\theta:=\arg u
$$
is therefore a physical parameter: it is the relative phase of the
$\Phi$-block noise coherence $u$ with respect to the Bell coherence. Explicitly, the
$|00\rangle$--$|11\rangle$ coherence of $\rho_\lambda$ is
$$
\langle00|\rho_\lambda|11\rangle=\tfrac{\lambda}{2}+(1-\lambda)|u|e^{i\theta},
$$
where the first term is the Bell-state contribution and the second is the
noise-state contribution. The choice of
$\PhiP$ as the reference Bell state is a convention, and the
analysis for any other Bell state is analogous. The endpoint phases
$\theta=0$ and $\theta=\pi$ are the cases in which the two coherences
add with the same sign or with opposite signs, which we call the
\emph{aligned} and \emph{anti-aligned} cases, respectively.
Normalization reduces the four populations to three independent parameters.
In the gauge $v\ge0$, the coherence sector is specified by the two magnitudes
$|u|$ and $v=|v|$ and the relative phase $\theta$. Thus, the complex $X$
family has six continuous threshold-relevant parameters, one more than the
real $X$ family.

\paragraph*{Singular-value flow.}
The CHSH-nonlocal threshold and the two- and three-setting CJWR
witness thresholds depend on the correlation tensor
$T_{ij}=\mathrm{Tr}[\rho\,(\sigma_i\otimes\sigma_j)]$.
The CHSH and CJWR criteria are known functions of the singular values of
$T$~\cite{HorodeckiCHSH1995,CostaAngelo2016}, which are local-unitary
invariants of the state. We therefore track these singular values as
$\lambda$ varies. Proposition~\ref{prop:flow} gives this dependence in
closed form.

\begin{proposition}[Singular-value flow]
\label{prop:flow}
In the gauge $v\ge0$ fixed above, the correlation tensor of
$\rho_\lambda$ is block diagonal for every complex $X$ state
{of Eq.~\eqref{eq:x-matrix}}. It contains a $2\times2$ block in the $xy$ plane
and the single entry $t_z$. Its three singular values are
$s_+(\lambda)$, $s_-(\lambda)$, and $|t_z(\lambda)|$, where the block
singular values are
\begin{align}
s_\pm(\lambda)&=\bigl|Q(\lambda)\pm P(\lambda)\bigr|,
\nonumber\\
Q(\lambda)&=\bigl|\lambda+2(1-\lambda)u\bigr|,
\qquad
P(\lambda)=2(1-\lambda)v,
\label{eq:sv-flow}
\end{align}
and the $z$ correlator depends affinely on $\lambda$,
\begin{equation}
t_z(\lambda)=\gamma+\lambda(1-\gamma),
\qquad
\gamma=a-b-c+d.
\label{eq:tz}
\end{equation}
\end{proposition}

The singular-value formulas are proved in Appendix~\ref{app:gauge}.
Together, Eqs.~\eqref{eq:sv-flow} and~\eqref{eq:tz} determine the
CHSH-nonlocal threshold and the CJWR witness thresholds. Their phase
dependence enters only through Eq.~\eqref{eq:sv-flow}. Inside the modulus
$Q=|\lambda+2(1-\lambda)u|$, the noise coherence adds to the
Bell weight as a complex amplitude. This addition gives
$Q^2=\lambda^2+4\lambda(1-\lambda)|u|\cos\theta+4(1-\lambda)^2|u|^2$,
whose cross term $4\lambda(1-\lambda)|u|\cos\theta$ carries the entire
phase dependence. The $\Psi$-block coherence $v$ only splits the
two $xy$ singular values symmetrically about $Q$, and its own phase has
already been gauged away. At the endpoints, Eq.~\eqref{eq:sv-flow}
returns the singular values of the noise state $\sigma_X$,
$s_\pm(0)=2\bigl||u|\pm v\bigr|$, and those of the Bell state,
$s_\pm(1)=1$, while Eq.~\eqref{eq:tz} gives the third singular value,
$|t_z(0)|=|\gamma|$ and $|t_z(1)|=1$. The flow interpolates between these
endpoint singular values as $\lambda$ runs from $0$ to $1$.

\paragraph*{Threshold conventions.}
For each operational ability $\mathcal O$ considered here, the Bell weights
at which $\rho_\lambda$ lacks $\mathcal O$ form a single interval. If this
interval is detached from $\lambda=0$, it has two boundary points, and a
convention is needed to specify which one is the threshold. The physical
reading fixes this choice. The intended state to be prepared and delivered is
$\PhiP$ at $\lambda=1$. Decreasing $\lambda$ adds noise weight and
moves the state toward $\sigma_X$. The upper boundary is therefore encountered
first and marks the loss of $\mathcal O$. If a lower boundary is present, it
marks a later recovery closer to the noise endpoint. We accordingly define the
\emph{definitive} threshold as the upper edge of the interval when it is
nonempty and as zero when it is empty,
\begin{equation}
\lambda_{\mathcal O}
:=\sup\{\lambda\in[0,1]:\rho_\lambda\text{ lacks }\mathcal O\},
\qquad
\sup\emptyset:=0.
\label{eq:definitive}
\end{equation}
The entanglement, teleportation usefulness, CJWR-$n$ witness,
CHSH-nonlocal, and $A\to B$ steerability thresholds are denoted by
$\lambda_*$, $\lambda_F$, $\lambda^{(n)}_{\rm CJWR}$,
$\lambda_{\rm CHSH}$, and $\lambda_{\rm steer}^{A\to B}$, respectively.

We give closed forms for the entanglement and teleportation
usefulness boundary crossings, valid whether $\sigma_X$ is
separable or entangled (Appendix~\ref{app:lower}). For separable
$\sigma_X$, they recover the thresholds of Ref.~\cite{eac-paper}, which
treats only that case. As functions of $\lambda$, the
optimized CHSH value and the optimized two- and three-setting CJWR witness
values are determined entirely by the singular values of
$T(\lambda)$~\cite{HorodeckiCHSH1995,CostaAngelo2016}. Consequently, the singular-value
flow determines the Bell weights at which each witness reaches its violation
bound.

\paragraph*{Real coherences.}
For $\theta\in\{0,\pi\}$, the coherence $u=\pm|u|$ is real and the
Pauli correlation tensors of the noise state and Bell state are
diagonal,
\begin{align}
T_{\sigma_X}&=\mathrm{diag}\bigl(2(u+v),\,2(v-u),\,a-b-c+d\bigr),
\nonumber\\
T_{\Phi^+}&=\mathrm{diag}(1,-1,1).
\label{eq:real-x-diagonal-tensor}
\end{align}
The singular values then become absolute values of affine functions.
Define
\begin{align}
t_x(\lambda)&=\alpha+\lambda(1-\alpha),
&
t_y(\lambda)&=\beta+\lambda(1-\beta),
\nonumber\\
\alpha&=2(u+v),
&
\beta&=2(u-v).
\label{eq:affine-correlators}
\end{align}
For convenience, $t_x(\lambda)=[T(\lambda)]_{xx}$,
$t_y(\lambda)=-[T(\lambda)]_{yy}$, and
$t_z(\lambda)=[T(\lambda)]_{zz}$. The sign convention for $t_y$ does not
affect the singular values.
The two block singular values are $\{s_+,s_-\}=\{|t_x|,|t_y|\}$.
Section~\ref{sec:alignment} proves that, with the populations
and coherence magnitudes fixed, the definitive thresholds are
nonincreasing in $\cos\theta$ (in Theorem~\ref{thm:mono}). Therefore, each
threshold reaches its minimum at the aligned
phase $\theta=0$ and its maximum at the anti-aligned phase $\theta=\pi$, as
stated explicitly in Corollary~\ref{cor:bounds}.

\begin{proposition}
\label{prop:windows}
The Bell weights $\lambda$ for which $\rho_\lambda$ is separable,
teleportation-useless, CJWR-$n$-satisfying ($n\in\{2,3\}$), or CHSH-local form
a single closed interval in each case, possibly empty. None of these intervals
contains $\lambda=1$. These intervals are nested as
$$
\begin{aligned}
\text{separable}
&\subseteq \text{teleportation-useless}
\\
&\subseteq \text{CJWR-$3$-satisfying}
\\
&\subseteq \text{CJWR-$2$-satisfying}=\text{CHSH-local}.
\end{aligned}
$$

The change of variable
$\eta=\lambda/(1-\lambda)$ defines a strictly increasing bijection from
$[0,1)$ onto $[0,\infty)$, with inverse
$\lambda=\eta/(1+\eta)$. It therefore preserves the ordering and interval
structure, and each finite boundary point in $\eta$ maps uniquely to a boundary
point in $\lambda$. The four intervals are as follows.

\begin{enumerate}
\item \emph{Separable.} Let
$D_{\rm sep}=bc-(\mathrm{Im}\,u)^2$. If $D_{\rm sep}<0$, no Bell weight gives
a separable state. For $D_{\rm sep}\ge0$, define
\begin{align}
\eta_\pm&=-2\,\mathrm{Re}\,u\pm2\sqrt{D_{\rm sep}},
\nonumber\\
\eta_\Psi&=\sqrt{(a-d)^2+4|v|^2}-(a+d).
\label{eq:sep-window}
\end{align}
The separable interval in $\eta$ is
$$
\bigl[\max\{0,\eta_\Psi,\eta_-\},\;\eta_+\bigr].
$$
This interval is empty when
$\max\{0,\eta_\Psi,\eta_-\}>\eta_+$.

\item \emph{Teleportation-useless.} Let
$D_F=(b+c)^2-4(\mathrm{Im}\,u)^2$. If $D_F<0$, no Bell weight gives a
teleportation-useless state. For $D_F\ge0$, define
\begin{align}
\eta_{F,\pm}&=-2\,\mathrm{Re}\,u\pm\sqrt{D_F},
\nonumber\\
\eta_{F,\Psi}&=2|v|-(a+d).
\label{eq:teleportation-window}
\end{align}
The teleportation-useless interval in $\eta$ is
$$
\bigl[\max\{0,\eta_{F,\Psi},\eta_{F,-}\},\;\eta_{F,+}\bigr].
$$
This interval is empty when
$\max\{0,\eta_{F,\Psi},\eta_{F,-}\}>\eta_{F,+}$.

\item \emph{CJWR-$3$-satisfying.} With $Q$, $P$, and $t_z$ from
Proposition~\ref{prop:flow}, this interval consists of the
$\lambda\in[0,1]$ satisfying
\begin{equation}
2\bigl(Q^2+P^2\bigr)+t_z^2\le1.
\label{eq:window-witness}
\end{equation}
{Its boundary points are roots of the quadratic equation given in
Appendix~\ref{app:chsh}.}

\item \emph{CHSH-local and CJWR-$2$-satisfying.} This interval consists of the
$\lambda\in[0,1]$ satisfying both
\begin{equation}
2\bigl(Q^2+P^2\bigr)\le1
\ \ \text{and}\ \ \bigl(Q+P\bigr)^2+t_z^2\le1.
\label{eq:window-chsh}
\end{equation}
Filtering the six possible roots of the quadratic and quartic equations in
Appendix~\ref{app:chsh} gives the boundary points.
\end{enumerate}
\end{proposition}

A proof of Proposition~\ref{prop:windows} is given in
Appendix~\ref{app:windows}. As a concrete example of detached intervals,
consider the noise state
$\sigma_X=\frac45|\Phi^-\rangle\!\langle\Phi^-|
+\frac15\,I_4/4$. The teleportation-useless interval coincides with the
separable interval, and the exact nesting is
$$
\left[\tfrac7{17},\tfrac9{19}\right]
\subset
\left[\tfrac{68-5\sqrt{35}}{163},\tfrac{68+5\sqrt{35}}{163}\right]
\subset
\left[\tfrac{32-15\sqrt{2}}{82},\tfrac{32+15\sqrt{2}}{82}\right].
$$

\begin{figure*}[!t]
\centering
\includegraphics{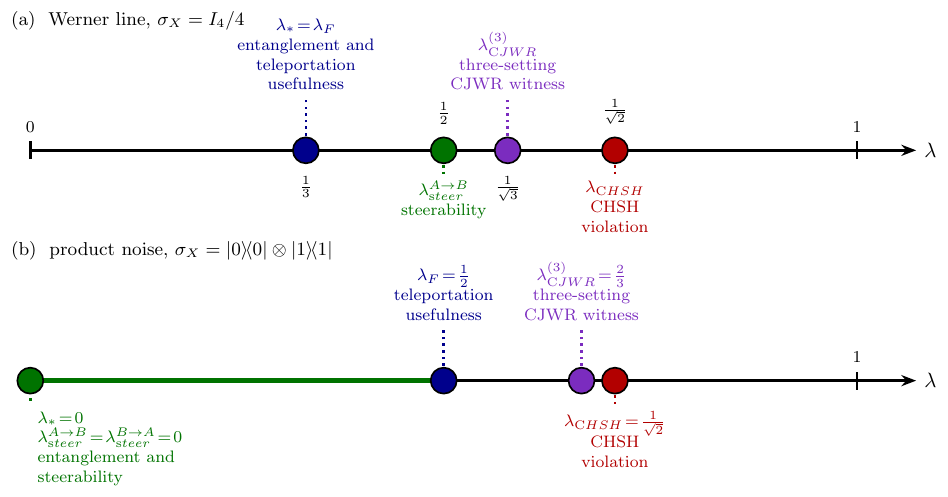}
\caption{Threshold order on two Bell-state mixtures.
{Panel~(a) shows the Bell weights at which the operational abilities
change along the Werner line $\sigma_X=I_4/4$.} The entanglement and
teleportation usefulness thresholds coincide at
$\lambda_*=\lambda_F=1/3$~\cite{Werner1989,HorodeckiTeleportation1999}. {The steerability threshold is
$1/2$ for both projective measurements~\cite{JonesWisemanDoherty2007}
and POVMs~\cite{ZhangChitambar2024}.} {The CJWR-3
witness threshold is $1/\sqrt3$, and CHSH violation occurs for
$\lambda>1/\sqrt2$. The boundary $\lambda_{\rm CHSH}=1/\sqrt2$ is also the
CJWR-2 witness threshold.}
{Panel~(b) shows the Bell weights at which the operational abilities
change for the mixture of $\PhiP$ with the product
state $|01\rangle\!\langle01|$ in
Eq.~\eqref{eq:steer-useless-line}.}
{For this mixture, $\lambda_*=0$, and
Theorem~\ref{cor:pure-noise-steering} gives the exact steerability thresholds
$\lambda_{\rm steer}^{A\to B}=\lambda_{\rm steer}^{B\to A}=0$,}
{whereas teleportation usefulness requires
$\lambda>\lambda_F=\tfrac12$.} {The thick green segment represents
$0<\lambda\le\tfrac12$, for which $\rho_\lambda$ is steerable in both
directions using projective measurements but teleportation-useless.}
{The CJWR-3 witness threshold is $2/3$, and CHSH violation occurs for
$\lambda>1/\sqrt2$ (Sec.~\ref{sec:true-steering}).}}
\label{fig:werner-hierarchy}
\end{figure*}

The state-set inclusion chain underlying Proposition~\ref{prop:windows} holds
in the full two-qubit state space. Because these sets are convex, the same
interval nesting holds along the Bell-mixing line for every two-qubit noise
state $\sigma$, whether or not $\sigma$ has $X$ form. In addition, the convex
set of states unsteerable from $A$ to $B$ under projective measurements gives
the chain
$$
\begin{aligned}
\text{separable}
&\subseteq \text{$A\to B$ unsteerable}
\\
&\subseteq \text{CJWR-$3$-satisfying}
\\
&\subseteq \text{CJWR-$2$-satisfying}=\text{CHSH-local}.
\end{aligned}
$$

\begin{corollary}
\label{cor:order}
For every two-qubit noise state $\sigma$, the definitive thresholds obey
\begin{align}
\lambda_*&\le\lambda_F\le\lambda^{(3)}_{\rm CJWR}
\le\lambda^{(2)}_{\rm CJWR}=\lambda_{\rm CHSH},
\label{eq:onset-order}
\\
\lambda_*&\le\lambda_{\rm steer}^{A\to B}\le\lambda^{(3)}_{\rm CJWR}
\le\lambda^{(2)}_{\rm CJWR}=\lambda_{\rm CHSH}.
\label{eq:witness-order}
\end{align}
The equality $\lambda^{(2)}_{\rm CJWR}=\lambda_{\rm CHSH}$ is proved in
Sec.~\ref{sec:cjwr-main}.
\end{corollary}

The two chains cannot be folded into a single chain because neither the teleportation-useless
set nor the set unsteerable from $A$ to $B$ under projective measurements
contains the other. Consequently, there is no universal order between
$\lambda_{\rm steer}^{A\to B}$ and $\lambda_F$. On the Werner line,
$\lambda_*=\lambda_F=1/3$, whereas
$\lambda_{\rm steer}^{A\to B}=1/2$ for both projective
measurements~\cite{JonesWisemanDoherty2007} {and
POVMs~\cite{ZhangChitambar2024}.} Thus, teleportation
usefulness does not imply steerability
[Fig.~\ref{fig:werner-hierarchy}(a)].
In the other direction, the Bell mixture with the product noise state
considered below has $\lambda_{\rm steer}^{A\to B}<\lambda_F$.

\paragraph*{Projective-measurement steerability without
teleportation usefulness.}
Consider the pure product noise
$\sigma_X=|0\rangle\!\langle0|\otimes|1\rangle\!\langle1|$. This is the
$X$ state with $b=1$ and all other entries equal to zero. Its mixture with
$\PhiP$ is
\begin{equation}
\rho_\lambda=
\begin{pmatrix}
\lambda/2&0&0&\lambda/2\\
0&1-\lambda&0&0\\
0&0&0&0\\
\lambda/2&0&0&\lambda/2
\end{pmatrix}.
\label{eq:steer-useless-line}
\end{equation}
For this case, Theorem~\ref{cor:pure-noise-steering} shows that $\rho_\lambda$
is steerable from $A$ to $B$ and from $B$ to $A$ if and only if it is
entangled. Together with the singular values of the correlation tensor, the
thresholds are
\begin{align*}
\lambda_*=\lambda_{\rm steer}^{A\to B}
=\lambda_{\rm steer}^{B\to A}&=0,
&
\lambda_F&=\tfrac12,\\
\lambda^{(3)}_{\rm CJWR}&=\tfrac23,
&
\lambda_{\rm CHSH}&=\tfrac1{\sqrt2}.
\end{align*}
At the endpoint $\lambda=0$, $\rho_0=|01\rangle\!\langle01|$ is separable
and unsteerable, whereas every $\rho_\lambda$ with $\lambda>0$ is steerable
in both directions using projective measurements. Hence, $\lambda=0$ is the
only unsteerable Bell weight. Standard teleportation usefulness requires
$f(\rho_\lambda)>1/2$~\cite{HorodeckiTeleportation1999}. Here,
$f(\rho_\lambda)=\max\{\lambda,(1-\lambda)/2\}$, so this condition holds
exactly when $\lambda>1/2$.
The exact
interval in which the states are steerable in both directions using projective
measurements but teleportation-useless is
$(0,\tfrac12]$ [Fig.~\ref{fig:werner-hierarchy}(b)].
{On the Werner line,
$\lambda_{\rm steer}^{A\to B}>\lambda_F$, whereas for the present
product-noise example, $\lambda_{\rm steer}^{A\to B}<\lambda_F$.
Thus, the relative order of the two thresholds depends on the noise state.}
{Section~\ref{sec:true-steering} compares the exact
steerability threshold with the CJWR witness thresholds.}

Figures~\ref{fig:sandwich}--\ref{fig:alignment} use separable noise, so at
$\lambda=0$ the state has none of the operational abilities considered here.
As the Bell weight increases, the abilities appear according to the applicable
threshold chains, with possible coincident thresholds. The
phase-diagram galleries in Figs.~\ref{fig:window-widening}
and~\ref{fig:gallery} provide a complementary view. They include noise states
that already possess some of these abilities at $\lambda=0$, up to CHSH
nonlocality, and display both the anchored and detached intervals described in
Proposition~\ref{prop:windows}.

\section{CHSH-nonlocal threshold}
\label{sec:chsh-main}

{Proposition~\ref{prop:flow} gives the singular values of the
correlation tensor of $\rho_\lambda$ in closed form. These singular values
determine the optimized CHSH value and hence the exact CHSH-nonlocal threshold
throughout the complex $X$ family. In the CHSH test, party $A$ measures one of the two
spin observables $\vec a_1\cdot\vec\sigma$ and
$\vec a_2\cdot\vec\sigma$, while party $B$ measures one of
$\vec b_1\cdot\vec\sigma$ and $\vec b_2\cdot\vec\sigma$. The four vectors are
unit vectors in $\mathbb R^3$ specifying the measurement directions, and
$\vec\sigma=(\sigma_x,\sigma_y,\sigma_z)$ is the vector of Pauli operators.
Throughout, arrows denote vectors. Each observable squares to the identity
and therefore defines a projective measurement with outcomes $\pm1$. The
corresponding CHSH operator is
\begin{equation}
\mathcal B_{\rm CHSH}
=\bigl(\vec a_1\cdot\vec\sigma\bigr)
 \otimes\bigl[(\vec b_1+\vec b_2)\cdot\vec\sigma\bigr]
+\bigl(\vec a_2\cdot\vec\sigma\bigr)
 \otimes\bigl[(\vec b_1-\vec b_2)\cdot\vec\sigma\bigr].
\label{eq:chsh-op}
\end{equation}
Every local-hidden-variable model satisfies
$|\langle\mathcal B_{\rm CHSH}\rangle_\rho|\le2$, where
$\langle\cdot\rangle_\rho=\mathrm{Tr}[\rho\,\cdot\,]$ denotes the expectation
value in the two-qubit state $\rho$~\cite{CHSH1969}.}
Because
$\langle\vec a\cdot\vec\sigma\otimes\vec b\cdot\vec\sigma\rangle_\rho
=\vec a^{\mathsf T}T\,\vec b$ for the correlation tensor $T$ of
Sec.~\ref{sec:xsetup}, the expectation of Eq.~\eqref{eq:chsh-op}
depends on the state only through $T$. Horodecki
{\textit{et al.} showed that the maximum over all four unit
measurement vectors is $2\sqrt{M(\rho)}$, where
$M(\rho)=s_{(1)}^2+s_{(2)}^2$ is the sum of the two largest squared singular
values of $T$~\cite{HorodeckiCHSH1995}. Thus, $\rho$
violates the CHSH inequality exactly when $M(\rho)>1$.} Along the line, we write
{$M(\lambda):=M(\rho_\lambda)$ and use the same convention for every
state function below. The CHSH-local interval is
$\{\lambda:M(\lambda)\le1\}$, and its definitive threshold
$\lambda_{\rm CHSH}$ is defined by Eq.~\eqref{eq:definitive}. We now derive
the interval edges.}

\begin{theorem}[CHSH-nonlocal threshold]
\label{thm:chsh}
{For the Bell mixture $\rho_\lambda$ of
Eq.~\eqref{eq:mixing-line}, where $\sigma_X$ is an arbitrary physical
two-qubit $X$ state with possibly complex coherences, the optimized CHSH
quantity is}
\begin{equation}
M(\lambda)=\max\bigl\{2\bigl[Q^2+P^2\bigr],\;
\bigl[Q+P\bigr]^2+t_z^2\bigr\}(\lambda).
\label{eq:chsh-max}
\end{equation}
{Here, $Q$, $P$, and $t_z$ are defined in
Proposition~\ref{prop:flow}.}
\end{theorem}

The maximum appears because $s_+\ge s_-$. The two largest singular
values are $s_+$ and the larger of $s_-$ and $|t_z|$.
The identity $s_\pm=|Q\pm P|$ turns the candidate sums
$s_+^2+s_-^2$ and $s_+^2+t_z^2$ into the two terms of
Eq.~\eqref{eq:chsh-max}. {The quadratic and quartic equations given in
Appendix~\ref{app:chsh} determine where the two branches of
Eq.~\eqref{eq:chsh-max} equal $1$. Each branch stays at or below $1$ on a single
interval of $\lambda$, and the CHSH-local interval is their intersection.
Filtering the six possible roots of these equations gives the boundary
points of the CHSH-local interval.} When the
coherences are real,
{$M=\max\{t_x^2+t_z^2,t_x^2+t_y^2,t_y^2+t_z^2\}$,} and,
whenever the CHSH-local {interval} is
nonempty, the definitive threshold reduces to
\begin{equation}
\lambda_{\rm CHSH}=\min\{R_{xz},R_{xy},R_{yz}\},
\label{eq:chsh-min-roots}
\end{equation}
where $R_{ij}$ is the largest root of the quadratic equation
$t_i(\lambda)^2+t_j(\lambda)^2=1$, and Appendix~\ref{app:chsh}
{gives both roots and their interval intersection.}

\section{CJWR witness thresholds}
\label{sec:cjwr-main}

The singular-value flow in Proposition~\ref{prop:flow} also
determines the optimized two- and three-setting CJWR witness thresholds and shows that
neither exceeds the CHSH-nonlocal threshold. Consider steering from $A$ to
$B$ for a two-qubit state. We use the $n$-setting CJWR test with
$n\in\{2,3\}$. For each setting $k=1,\dots,n$, party $A$ reports an outcome
$A_k=\pm1$, while party $B$ measures $\vec b_k\cdot\vec\sigma$, where the
directions $\vec b_k\in\mathbb R^3$ are mutually orthogonal unit vectors. For
every local-hidden-state model, the resulting correlations satisfy the CJWR
inequality~\cite{CavalcantiJonesWisemanReid2009}
\begin{equation}
\frac{1}{\sqrt n}\Bigl|\sum_{k=1}^{n}
\langle A_k\,\vec b_k\!\cdot\!\vec\sigma\rangle_\rho\Bigr|\le1,
\label{eq:cjwr-ineq}
\end{equation}
so a violation certifies steering from $A$ to $B$. Let
$s_{(1)}\ge s_{(2)}\ge s_{(3)}$ be the singular values of $T$. Maximizing the
left-hand side of Eq.~\eqref{eq:cjwr-ineq} over all allowed measurement
settings gives
\begin{equation}
F_n(\rho)=\Bigl(\textstyle\sum_{i=1}^{n}s_{(i)}^2\Bigr)^{1/2},
\label{eq:cjwr-F-def}
\end{equation}
as shown by Costa and Angelo~\cite{CostaAngelo2016}. Steering is
certified when $F_n>1$. Because $F_n\le1$ does not establish unsteerability
under all projective measurements,
$\lambda_{\rm steer}^{A\to B}\le\lambda^{(n)}_{\rm CJWR}$. Two identities
follow immediately. First, $F_2^2=s_{(1)}^2+s_{(2)}^2=M$, so the
two-setting CJWR inequality and the optimized CHSH inequality are
violated by exactly the same states. Second,
$F_3^2=M+s_{(3)}^2\ge F_2^2$, so the three-setting witness certifies
steering wherever the two-setting one does.

\begin{theorem}[CJWR witness thresholds]
\label{thm:cjwr}
For the Bell mixture $\rho_\lambda$ of
Eq.~\eqref{eq:mixing-line}, where $\sigma_X$ is an arbitrary physical
two-qubit $X$ state with possibly complex coherences, the optimized CJWR
values satisfy
\begin{align}
F_3(\lambda)^2&=2\bigl[Q^2+P^2\bigr](\lambda)+t_z(\lambda)^2,
\nonumber\\
F_2(\lambda)^2&=M(\lambda).
\label{eq:cjwr-F}
\end{align}
\end{theorem}

Two consequences follow. First,
$\lambda^{(2)}_{\rm CJWR}=\lambda_{\rm CHSH}$ identically. Second,
$F_3^2$ is a quadratic polynomial in $\lambda$, because $Q^2$,
$P^2$, and $t_z^2$ are quadratic. When the equation
$F_3(\lambda)^2=1$ has real roots $R_3^-\le R_3^+$, the
CJWR-$3$-satisfying set is $[R_3^-,R_3^+]\cap[0,1]$. This set may be empty
and reduces to a single point at a double root. The definitive
$\lambda^{(3)}_{\rm CJWR}$ of Eq.~\eqref{eq:definitive} is $R_3^+$ when
the intersection is nonempty and is zero otherwise. If the equation has no
real root, the set is empty and $\lambda^{(3)}_{\rm CJWR}=0$. The determination of the
CJWR-$3$-satisfying boundaries is therefore much simpler than for
the CHSH-local case. When the coherences are real,
$\lambda^{(3)}_{\rm CJWR}$ is the largest root of the
quadratic equation
$t_x(\lambda)^2+t_y(\lambda)^2+t_z(\lambda)^2=1$, and
Appendix~\ref{app:chsh} records both
roots in closed form. The inequality
$F_3\ge F_2$ and the identity
$F_2^2=M$ supply the CJWR
nesting that Proposition~\ref{prop:windows} uses and complete the
proof of Corollary~\ref{cor:order}.

\section{Exact and finite-setting steerability thresholds}
\label{sec:true-steering}

Exact steerability thresholds are substantially harder to determine than the
witness thresholds of the previous two sections. A witness threshold needs one
violated inequality, whereas an exact threshold must rule out every local
hidden state model for the full measurement class, so the CJWR values of
Sec.~\ref{sec:cjwr-main} bound
$\lambda_{\rm steer}^{A\to B}$ from above without determining it. Apart
from the Werner line, where projective measurements give
$\lambda_{\rm steer}^{A\to B}=1/2$~\cite{JonesWisemanDoherty2007}, exact values
on the mixing line are scarce. We obtain exact thresholds in two
rank-deficient sectors, Bell mixtures with an arbitrary pure noise state and
{two diagonal families of mixed $X$ noise with one missing
population.} In those cases, the
steerability threshold collapses onto the entanglement threshold, and
arbitrary POVMs do not lower it. For generic full-rank mixed $X$ noise, we
instead report finite-setting upper bounds. These bounds come from
a semidefinite program that tests local-hidden-state models for a fixed list
of $N$ measurement directions. We evaluate it for two nested lists of
eight and sixteen directions and compare the results with the CJWR witness thresholds of
Sec.~\ref{sec:cjwr-main}. Denote the $A\to B$
steerability threshold under arbitrary POVMs by
$\lambda_{\rm steer,POVM}^{A\to B}$. The $B\to A$ threshold is denoted
analogously.

The exact results rest on three recent theorems, quoted here in the form
in which they are used. The effective local dimension of a party is the
rank of its reduced state.

\begin{lemma}[Rank-two steerability~\cite{RankTwoProjectiveSteering2026}]
\label{lem:rank-two}
Every rank-two bipartite entangled state in arbitrary finite local dimensions
is steerable in at least one direction using projective measurements. When the
two effective local dimensions are equal, it is steerable in both directions
using projective measurements.
\end{lemma}

\begin{lemma}[Pure conditional state~\cite{PureSteeredState2026}]
\label{lem:pure-steered}
Let $\rho_{AB}$ be an entangled two-qubit state. If some projective
measurement on either party has a nonzero-probability outcome whose
normalized conditional state on the other party is pure, then
$\rho_{AB}$ is steerable in both directions using projective
measurements.
\end{lemma}

\begin{lemma}[Product vector in the kernel~\cite{ProductNullSteering2026}]
\label{lem:product-null}
Let $\rho$ be a two-qubit state satisfying
$$
\rho\,|e,f\rangle=0
$$
for a product vector $|e,f\rangle=|e\rangle\otimes|f\rangle$. Let
$|e^\perp\rangle$ and $|f^\perp\rangle$ complete the corresponding local
orthonormal bases, and define
$$
h=\langle e^\perp\!,f|\,\rho\,|e,f^\perp\rangle .
$$
Then $\rho$ is entangled if and only if $h\ne0$. In that case, the principal
submatrix of $\rho^{T_B}$ on
$$
\operatorname{span}\{|e,f\rangle,\,|e^\perp\!,f^\perp\rangle\}
$$
has determinant $-|h|^2<0$, and $\rho$ is steerable in both directions using
projective measurements.
\end{lemma}

{Lemmas~\ref{lem:rank-two} and~\ref{lem:pure-steered} apply to
mixtures with pure noise. Lemma~\ref{lem:pure-steered} covers the case in which
$\rho_\lambda$ has rank one, and Lemma~\ref{lem:rank-two} covers the case in
which it has rank two. For the two diagonal families below,
Lemmas~\ref{lem:pure-steered} and~\ref{lem:product-null} provide independent
proofs of steerability in both directions using projective measurements. We
use Lemma~\ref{lem:product-null} because its hypotheses can be checked directly
from the matrix of $\rho_\lambda$. We then exhibit the pure conditional state
required by Lemma~\ref{lem:pure-steered}.}

{
\paragraph*{Proof roadmap.}
Every separable state is unsteerable under arbitrary POVMs, and every
projective measurement is a POVM~\cite{WisemanJonesDoherty2007}. The
definitive thresholds therefore obey
$$
\lambda_*\le\lambda_{\rm steer,POVM}^{A\to B}
\le\lambda_{\rm steer}^{A\to B},
$$
and the same inequalities hold from $B$ to $A$. {If, for every
$\lambda$, entanglement is equivalent to projective-measurement steerability
in both directions, then the same equivalence holds for steering with
arbitrary POVMs. All four steerability thresholds then equal $\lambda_*$.}
The proofs below therefore only need to show that every entangled
$\rho_\lambda$ is steerable in both directions using projective measurements.
}

\subsection{Arbitrary pure noise}

\begin{theorem}
\label{cor:pure-noise-steering}
{Let $\sigma=|\psi\rangle\!\langle\psi|$ be any pure two-qubit
noise state, and let
$$
\rho_\lambda
=\lambda\PhiP
+(1-\lambda)|\psi\rangle\!\langle\psi|.
$$
For every $\lambda\in[0,1]$, the state $\rho_\lambda$ is entangled if and only
if it is steerable in both directions using projective measurements. The same
equivalence holds for steering with arbitrary POVMs. Thus, POVMs offer no
advantage over projective measurements for steering in either direction along
this mixture, and
$$
\lambda_*
=\lambda_{\rm steer}^{A\to B}
=\lambda_{\rm steer,POVM}^{A\to B}
=\lambda_{\rm steer}^{B\to A}
=\lambda_{\rm steer,POVM}^{B\to A}.
$$
}
\end{theorem}

\paragraph*{Proof.}
{By the proof roadmap above, only one implication needs proof,
namely that an entangled $\rho_\lambda$ is steerable in both directions using
projective measurements.

Fix $\lambda\in[0,1]$ and let
$S=\operatorname{span}\{|\Phi^+\rangle,|\psi\rangle\}$. Every
$|u\rangle\in S^\perp$ is annihilated by $\rho_\lambda$, and
$\dim S^\perp\ge2$ in the four-dimensional two-qubit space, so $\rho_\lambda$
has rank one or rank two.

{If $\rho_\lambda$ has rank one, then, since it is entangled,
both Schmidt coefficients are nonzero. A Schmidt-basis measurement
on either party therefore prepares a pure conditional state on the other party
with nonzero probability. Lemma~\ref{lem:pure-steered} gives steerability in
both directions using projective measurements.}

If the rank is two, then $0<\lambda<1$, because $\rho_0$ and $\rho_1$ are
pure. The reduced states of $\PhiP$ are maximally
mixed, so
$$
(\rho_\lambda)_A=\frac{\lambda}{2}I+(1-\lambda)\psi_A,
\qquad
(\rho_\lambda)_B=\frac{\lambda}{2}I+(1-\lambda)\psi_B,
$$
where $\psi_A$ and $\psi_B$ are the positive semidefinite reduced states of
$|\psi\rangle\langle\psi|$. Both marginals are then bounded below by
$(\lambda/2)I$, hence positive definite and of rank two. The two effective
local dimensions are therefore both equal to two, and
Lemma~\ref{lem:rank-two} gives steerability in both directions using
projective measurements. {Thus, for every $\lambda$,
$\rho_\lambda$ is entangled if and only if it is steerable in both directions
using projective measurements. Since projective measurements are POVMs and
steering implies entanglement, $\rho_\lambda$ is entangled if and only if it
is steerable in both directions when arbitrary POVMs are allowed. Hence, the
entanglement threshold and the four steerability thresholds in
Theorem~\ref{cor:pure-noise-steering} are equal.}}
\hfill$\square$

{\paragraph*{Examples of pure-noise thresholds.}
The common threshold in Theorem~\ref{cor:pure-noise-steering} can be zero or
positive. Because the theorem identifies the separable and unsteerable Bell
weights for both measurement classes and both directions, it is enough to
determine the separable Bell weights in each example.
The product-noise mixture with $|\psi\rangle=|01\rangle$ provides the
zero-threshold example discussed in Sec.~\ref{sec:xsetup} and
Eq.~\eqref{eq:steer-useless-line}.
A positive common threshold occurs for $|\psi\rangle=|\Phi^-\rangle$. The four
eigenvalues of $\rho_\lambda^{T_B}$ are $1/2$, $1/2$, and
$\pm(2\lambda-1)/2$. Thus, $\rho_\lambda$ is entangled unless
$\lambda=1/2$. At that Bell weight,
$\rho_{1/2}=(|00\rangle\langle00|+|11\rangle\langle11|)/2$ is separable. The
separable and the unsteerable Bell weights are therefore $\{1/2\}$, and all
five thresholds equal $1/2$. Here, $\lambda=1/2$ is an isolated boundary
contact rather than an onset.}

\subsection{\texorpdfstring{Diagonal $X$ noise with one missing
population}{Diagonal X noise with one missing population}}

{The next result treats two diagonal families of mixed $X$ noise that lie on
the boundary of state space and carry a product vector in the kernel.}

\begin{proposition}
\label{prop:boundary-steering}
{Let $a\ge0$, $b\ge0$, and $a+b\le1$, and let the noise state be
$$
\sigma_{a,b}=a|00\rangle\!\langle00|+b|01\rangle\!\langle01|
+(1-a-b)|11\rangle\!\langle11|.
$$
Define
$\rho_\lambda=\lambda\PhiP+(1-\lambda)\sigma_{a,b}$.
For every $\lambda\in(0,1]$, the state $\rho_\lambda$ is entangled and
steerable in both directions using projective measurements. Consequently,
$$
\lambda_*=\lambda_{\rm steer}^{A\to B}=\lambda_{\rm steer}^{B\to A}=0.
$$
The teleportation usefulness threshold, the two CJWR witness thresholds, and
the CHSH-nonlocal threshold depend only on $b$:
\begin{align}
\lambda_F&=\frac{b}{1+b},
\label{eq:boundary-lower}\\
\lambda^{(3)}_{\rm CJWR}
&=\frac{b(2b-1)+\sqrt{b(2-b)}}{1+2b^2},
\nonumber\\
\lambda^{(2)}_{\rm CJWR}=\lambda_{\rm CHSH}
&=\min\!\left\{\frac{1}{\sqrt2},\,
\frac{2[b(2b-1)+\sqrt b]}{1+4b^2}\right\}.
\label{eq:boundary-upper}
\end{align}
The Bell weights for which $\rho_\lambda$ is steerable in both directions
using projective measurements but useless for standard teleportation form the
interval $(0,b/(1+b)]$. This interval is empty when $b=0$. Replacing
$|01\rangle$ by $|10\rangle$ exchanges the two qubits and leaves all these
thresholds unchanged.}
\end{proposition}

\paragraph*{Proof.}
{{In the computational basis,
$$
\rho_\lambda=
\begin{pmatrix}
A_\lambda&0&0&\lambda/2\\
0&(1-\lambda)b&0&0\\
0&0&0&0\\
\lambda/2&0&0&D_\lambda
\end{pmatrix}.
$$
Here, $A_\lambda=\lambda/2+(1-\lambda)a$ and
$D_\lambda=\lambda/2+(1-\lambda)(1-a-b)$.
After partial transposition, the principal block on
$\operatorname{span}\{|01\rangle,|10\rangle\}$ has determinant
$-\lambda^2/4$. Thus, the partial transpose is not
positive semidefinite for any $\lambda\in(0,1]$, and $\rho_\lambda$ is
entangled~\cite{Peres1996,Horodecki1996}. At $\lambda=0$,
$\rho_0=\sigma_{a,b}$ is diagonal in a product basis and is separable.
Therefore, $\lambda_*=0$.}

Neither $\PhiP$ nor
$\sigma_{a,b}$ has population in $|10\rangle$, so
$|10\rangle\in\ker\rho_\lambda$. In Lemma~\ref{lem:product-null}, take
$|e\rangle=|1\rangle$, $|f\rangle=|0\rangle$,
$|e^\perp\rangle=|0\rangle$, and $|f^\perp\rangle=|1\rangle$. The coupling is
$$
h=\langle00|\rho_\lambda|11\rangle=\frac{\lambda}{2}.
$$
Under partial transpose, this coherence becomes the off-diagonal entry of
the principal block displayed above, whose determinant is
$-|h|^2=-\lambda^2/4$. Lemma~\ref{lem:product-null} therefore gives
steerability in both directions using projective measurements for every
$\lambda\in(0,1]$, and both steerability thresholds are zero.

Lemma~\ref{lem:pure-steered} corroborates this conclusion through
explicit pure conditional states in both directions. For
$\lambda\in(0,1]$,
applying the projector $|1\rangle\!\langle1|$ on $A$ prepares the unnormalized
conditional
state
$$
\left[\frac{\lambda}{2}+(1-\lambda)(1-a-b)\right]
|1\rangle\!\langle1|
$$
on $B$, and applying the projector $|0\rangle\!\langle0|$ on $B$ prepares the
unnormalized conditional state
$$
\left[\frac{\lambda}{2}+(1-\lambda)a\right]
|0\rangle\!\langle0|
$$
on $A$. Both probabilities are at least $\lambda/2>0$, and both
normalized conditional states are pure. For each $\lambda\in(0,1]$,
$\rho_\lambda$ is entangled by the partial-transpose argument above and admits
either displayed nonzero-probability pure conditional state.
Lemma~\ref{lem:pure-steered} therefore gives steerability in both directions.
Displaying both outcomes makes the two steering directions explicit.

The correlation tensor determines all remaining thresholds. For this family,
$$
T(\rho_\lambda)
=\operatorname{diag}\!\left(
\lambda,-\lambda,1-2b(1-\lambda)\right).
$$
The corresponding fully entangled fraction is
$$
f(\rho_\lambda)
=\frac12\max\!\left\{
1-b+(1+b)\lambda,\;b(1-\lambda)\right\}.
$$
The parameter $a$ moves weight between $|00\rangle$ and $|11\rangle$ and
therefore changes the local marginals. However, it leaves the sum of these two
populations and the coherence unchanged. Consequently, the correlation
tensor, the fully entangled fraction, and all four remaining thresholds depend
only on $b$.

Standard teleportation is useful exactly when $f(\rho_\lambda)>1/2$. In the
expression for $f(\rho_\lambda)$ above, the second branch
$b(1-\lambda)$ never exceeds $1$. Therefore, $f(\rho_\lambda)>1/2$ holds
exactly when the first branch satisfies
$1-b+(1+b)\lambda>1$. This condition is equivalent to
$\lambda>b/(1+b)$ and gives the teleportation usefulness threshold in
Eq.~\eqref{eq:boundary-lower}.

For this family, the longitudinal correlation is
$t_z(\lambda)=1-2b(1-\lambda)$. The three singular values of the correlation
tensor are therefore $\lambda$, $\lambda$, and $|t_z(\lambda)|$. Hence, the
optimized three-setting CJWR value and the optimized CHSH quantity reduce to
\begin{align*}
F_3(\lambda)^2&=2\lambda^2+t_z(\lambda)^2,\\
M(\lambda)&=\max\{2\lambda^2,\lambda^2+t_z(\lambda)^2\}.
\end{align*}
The state is CJWR-$3$-satisfying exactly when $F_3(\lambda)^2\le1$. At
$\lambda=0$, this quantity is $(1-2b)^2\le1$, and at $\lambda=1$, it is
$3>1$. Because $F_3(\lambda)^2$ is a convex quadratic, the upper edge of the
CJWR-$3$-satisfying interval is the larger root of $F_3(\lambda)^2=1$.
Solving this equation gives the stated CJWR-$3$ witness threshold.

The CHSH-local condition $M(\lambda)\le1$ reduces to the simultaneous
inequalities
$$
2\lambda^2\le1,
\qquad
\lambda^2+t_z(\lambda)^2\le1.
$$
The first inequality holds up to $\lambda=1/\sqrt2$. The second is a convex
quadratic inequality that holds at $\lambda=0$, so its upper boundary is the
larger root of $\lambda^2+t_z(\lambda)^2=1$. The smaller of these two boundary
values gives the CHSH-nonlocal threshold in Eq.~\eqref{eq:boundary-upper}.

Intersecting the projective-measurement steerable set $(0,1]$ with the
teleportation-useless interval $[0,b/(1+b)]$ gives the interval in the
proposition. Finally, the swap operator maps $|01\rangle$ to $|10\rangle$,
exchanges the two steering directions, and leaves the fully entangled fraction
and the singular values of the correlation tensor unchanged. Hence, all
displayed thresholds are unchanged under this replacement.}
\hfill$\square$

\begin{corollary}
\label{cor:povm-projective}
{Consider a noise state $\sigma_X$ of either diagonal form
$\operatorname{diag}(a,b,0,d)$ or $\operatorname{diag}(a,0,c,d)$. The entries
are nonnegative and satisfy $a+b+d=1$ or $a+c+d=1$, respectively. The
corresponding Bell mixture is
$$
\rho_\lambda
=\lambda\PhiP+(1-\lambda)\sigma_X.
$$
For every $\lambda\in[0,1]$, the state $\rho_\lambda$ is entangled if and only
if it is steerable in both directions using projective measurements. The same
equivalence holds for steering with arbitrary POVMs. Moreover,
$\rho_\lambda$ is entangled if and only if $\lambda>0$. Thus, POVMs offer no
advantage over projective measurements for steering in either direction along
those mixtures, and
$$
\lambda_*
=\lambda_{\rm steer}^{A\to B}
=\lambda_{\rm steer,POVM}^{A\to B}
=\lambda_{\rm steer}^{B\to A}
=\lambda_{\rm steer,POVM}^{B\to A}=0.
$$
}
\end{corollary}

This corollary restates
Proposition~\ref{prop:boundary-steering} for the two orderings of the missing
population, and the POVM statement follows from the proof roadmap.

\subsection{Finite-setting upper bounds for generic mixed
\texorpdfstring{$X$}{X} noise}

{The preceding exact arguments cover arbitrary pure noise and diagonal
$X$ noise with one missing population. They do not determine the exact
steerability threshold for generic full-rank mixed $X$ noise. For each fixed
noise state $\sigma_X$ and fixed set
$\mathcal M_N=\{\vec n_k\}_{k=1}^N$ of $N$ projective measurement directions
on $A$, we construct a semidefinite program (SDP) to find the largest Bell
weight for which the conditional states on $B$ produced by those measurements
can be reproduced by a single LHS model, subject to the constraints established
by the standard finite-assemblage LHS
formulation~\cite{SkrzypczykNavascuesCavalcanti2014,
CavalcantiSkrzypczyk2017}. Its exact optimum is denoted by
$\lambda^{(N)}_{\rm det}$. Different fixed choices of $\mathcal M_N$ can give
different finite-setting upper bounds. For the reported comparison, we use
the same fixed nested sets $\mathcal M_8\subset\mathcal M_{16}$ for every
noise state.

The procedure has three steps. First, fix the direction set
$\mathcal M_N=\{\vec n_k\}_{k=1}^N$ of unit vectors on the Bloch sphere. Each
vector specifies a binary projective measurement on $A$,
$$
\Pi_{a|k}=\frac{I+a\vec n_k\cdot\vec\sigma}{2},
\qquad a\in\{+1,-1\}.
$$
Second, at Bell weight $\lambda$, apply every selected measurement to
$\rho_\lambda$ and calculate the unnormalized conditional states on $B$,
\begin{equation}
\sigma_{a|k}(\lambda)
=\operatorname{Tr}_A\!\left[(\Pi_{a|k}\otimes I)\rho_\lambda\right]
\label{eq:finite-assemblage}
\end{equation}
for every $a$ and $k$. These $2N$ operators form the finite assemblage. The
trace of $\sigma_{a|k}$ is the probability of outcome $a$, and the normalized
operator $\sigma_{a|k}/\operatorname{Tr}\sigma_{a|k}$ is the conditional
state on $B$ when this probability is nonzero.

Third, test whether one common hidden-state ensemble can reproduce all $2N$
assemblage elements simultaneously. At fixed $\lambda$, feasibility means
that the selected measurements admit an LHS model, while infeasibility
certifies steering with those measurements. The following program performs
this test while maximizing $\lambda$:
\begin{equation}
\begin{aligned}
\lambda^{(N)}_{\rm det}
&={\max_{\lambda,\{\omega_\xi\}}\lambda}\\
&\text{subject to}\quad 0\le\lambda\le1,\\
&{\omega_\xi\succeq0,\quad \xi\in\{+1,-1\}^N,}\\
&\sum_\xi\omega_\xi=\rho_B(\lambda),\\
&\sum_{\xi:\,\xi_k=+1}\omega_\xi
=\sigma_{+1|k}(\lambda),\quad k=1,\ldots,N,
\end{aligned}
\label{eq:finite-lhs-program}
\end{equation}
Here, $0\le\lambda\le1$ is the physical range. The remaining constraints
encode an LHS model for the finite assemblage in the variables
$\omega_\xi$. The string $\xi=(\xi_1,\ldots,\xi_N)\in\{+1,-1\}^N$
is a response table with one preassigned outcome $\xi_k$ for each setting
$k$. Here, $p_\xi=\operatorname{Tr}\omega_\xi$ is its probability
and, for $p_\xi>0$, $\tau_\xi=\omega_\xi/p_\xi$ is the corresponding hidden
state. The variables describe a classical prepare-and-send simulation. Before
a setting is chosen, a source samples the response table $\xi$ with probability
$p_\xi$, gives the table to $A$, and sends the qubit state $\tau_\xi$ to $B$.
When $A$ is asked for the outcome of setting $k$, $A$ reads the $k$th entry and
announces $a=\xi_k$. Thus, the outcome announced by $A$ is fixed by the sampled
table, while the state received by $B$ was prepared before $k$ was chosen and
is independent of $k$. The first
constraint requires the average prepared state to reproduce the reduced state of $B$,
$\rho_B(\lambda)=\operatorname{Tr}_A\rho_\lambda$, and the constraint for
setting $k$ requires the runs in which measurement setting $k$ is used and
outcome $+1$ is announced to build up the conditional state
$\sigma_{+1|k}(\lambda)$. Feasibility at $\lambda$
therefore means that the selected assemblage admits this LHS simulation and
does not demonstrate steering. Appendix~\ref{app:numerics} derives this
encoding from the general LHS model.

Both $\rho_B(\lambda)$ and the conditional states in
Eq.~\eqref{eq:finite-assemblage} are affine in $\lambda$. Therefore,
Eq.~\eqref{eq:finite-lhs-program} maximizes $\lambda$ while searching over the
hidden-state operators $\{\omega_\xi\}$ that satisfy the LHS constraints. No
scan over fixed Bell weights is required. For fixed $\mathcal M_N$, closedness
and convexity of the finite-assemblage LHS set make the feasible Bell weights a
closed interval, possibly empty. The convention of Eq.~\eqref{eq:definitive}
defines $\lambda^{(N)}_{\rm det}$ as the definitive upper edge of this interval
or as zero if the interval is empty. Because the feasible interval
is a closed subset of $[0,1]$, its upper edge is attained whenever the
interval is nonempty. For every numerical family below, $\lambda=0$ is
feasible because $\sigma_X$ is separable, so the maximum in
Eq.~\eqref{eq:finite-lhs-program} is attained. For a noise state whose
finite-setting feasible set is empty, the convention
$\lambda^{(N)}_{\rm det}=0$ applies instead.

The finite-setting optimum and the optimized CJWR witness threshold give two
upper bounds on the projective-measurement steerability threshold:
\begin{equation}
\lambda_{\rm steer}^{A\to B}\le\lambda^{(N)}_{\rm det},
\qquad
\lambda_{\rm steer}^{A\to B}\le\lambda^{(n)}_{\rm CJWR}.
\label{eq:enclosure}
\end{equation}
The two bounds need not be ordered. For the nested pair
$\mathcal M_8\subset\mathcal M_{16}$, set inclusion guarantees
$\lambda^{(16)}_{\rm det}\le\lambda^{(8)}_{\rm det}$ pointwise. The quantity
$\lambda^{(N)}_{\rm det}$ is the largest Bell-state coefficient $\lambda$ in
Eq.~\eqref{eq:mixing-line} for which the selected $N$ measurements admit an
LHS model. Its solver estimate is $\widehat{\lambda}^{(N)}_{\rm det}$.
Appendix~\ref{app:numerics} specifies the direction sets and solver settings.
At $\lambda=\widehat{\lambda}^{(N)}_{\rm det}$, the LHS
constraints have a numerical solution, and Appendix~\ref{app:numerics} reports
floating-point constraint errors below $7\times10^{-10}$.}

The program does not rely on the $X$-state structure. For any two-qubit noise
state, it constructs the finite assemblage of
Eq.~\eqref{eq:finite-assemblage} for the corresponding Bell mixture and thereby
gives an upper bound on its directional projective-measurement steerability
threshold. We apply it below to three separable $X$-noise families, including
one diagonal product family.

\begin{figure}[!t]
\centering
\includegraphics[width=\columnwidth]{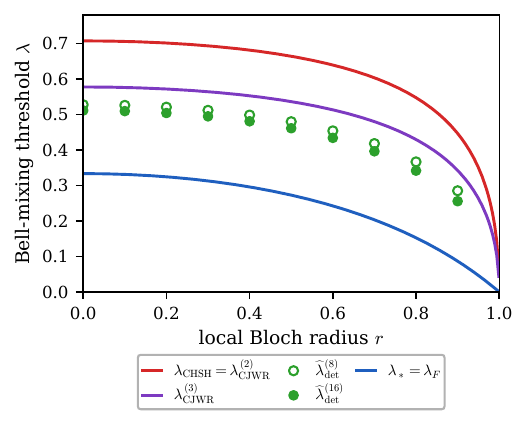}
\caption{{Operational thresholds on the diagonal $X$ family}
$\sigma_X=\sigma_A\otimes\sigma_B$ with
$\sigma_A=\sigma_B=(I+r\sigma_z)/2$, plotted against the local Bloch-vector
length $r$. The curves show the CHSH-nonlocal threshold, which is also the
CJWR-$2$ witness threshold, the CJWR-$3$ witness threshold, and
$\lambda_*=\lambda_F$.
Green markers show $\widehat{\lambda}^{(8)}_{\rm det}$ and
$\widehat{\lambda}^{(16)}_{\rm det}$, the numerical estimates of the
two finite-setting thresholds. The corresponding exact finite-setting optima bound the
projective-measurement steerability threshold from above. At $r=0$, the eight-
and sixteen-direction estimates are $0.527$ and $0.511$,
respectively, compared with the exact Werner value
$\lambda_{\rm steer}^{A\to B}=1/2$~\cite{JonesWisemanDoherty2007}.}
\label{fig:sandwich}
\end{figure}

\begin{figure*}[!t]
\centering
\includegraphics[width=0.92\textwidth]{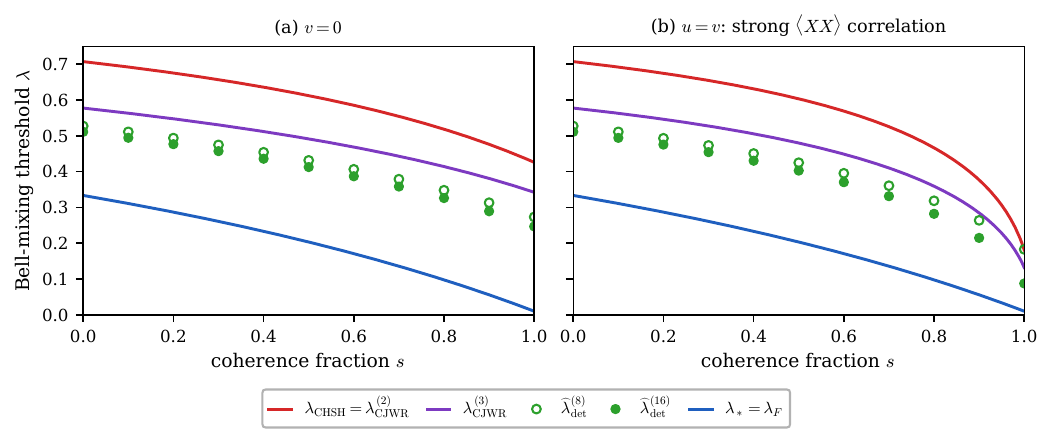}
\caption{{Operational thresholds on two families of separable $X$ states
with diagonal $(0.30,0.25,0.25,0.20)$ and real coherence scaled by a
fraction $s$ up to the separability boundary. Panel~(a) has $v=0$,
and panel~(b) has $u=v$. On both families, the definitive thresholds shown
by the solid curves have the order in Corollary~\ref{cor:order}. Green markers
show the numerical estimates
$\widehat{\lambda}^{(8)}_{\rm det}$ and
$\widehat{\lambda}^{(16)}_{\rm det}$. The corresponding exact
finite-setting optima bound the projective-measurement steerability threshold
from above. The
sixteen-direction estimate lies below the CJWR-$3$ witness
threshold, which lies below
$\lambda_{\rm CHSH}=\lambda^{(2)}_{\rm CJWR}$ (red). Thus, the
numerical comparison indicates finite-setting steering detection below the
optimized CJWR-$3$ witness threshold.}}
\label{fig:gap}
\end{figure*}

{\paragraph*{Numerical test families.}
We evaluate the diagonal product family in Fig.~\ref{fig:sandwich} and two
real-coherence families in the two panels of Fig.~\ref{fig:gap}. The product
family is
$$
\sigma_{\rm prod}(r)
=\frac{I+r\sigma_z}{2}\otimes\frac{I+r\sigma_z}{2},
\qquad 0\le r\le1.
$$
The two real-coherence families share the fixed diagonal
$(a,b,c,d)=(0.30,0.25,0.25,0.20)$ and have
$(u,v)=(s\sqrt{ad},0)$ and $(u,v)=(s\sqrt{ad},s\sqrt{ad})$, respectively,
where $s\in[0,1]$ is the coherence fraction. In both families, $|u|$ and
$|v|$ are at most $\sqrt{ad}$.
Hence, $\sigma_X$ and its partial transpose are positive semidefinite, so these
two-qubit noise states are separable. The product family is separable by
construction. Every family is invariant under exchanging $A$ and $B$, so the
two directional finite-setting thresholds coincide.}

{\paragraph*{Nested refinement.}
Across all $32$ sampled points of Figs.~\ref{fig:sandwich}
and~\ref{fig:gap}, the observed refinement
$\widehat{\lambda}^{(8)}_{\rm det}-
\widehat{\lambda}^{(16)}_{\rm det}$ ranges from $0.0162$ to $0.0953$. The
largest change occurs at $s=1$, the endpoint of the $u=v$ family, where
$\widehat{\lambda}^{(8)}_{\rm det}=0.1827$ drops to
$\widehat{\lambda}^{(16)}_{\rm det}=0.0874$. At this point, the CJWR-$3$
witness threshold is $0.132$, below the eight-setting estimate and above the
sixteen-setting estimate. The optimized CJWR-$3$ witness therefore gives a
tighter bound than $\mathcal M_8$. This does not contradict measurement-set
inclusion because the CJWR-$3$ witness is optimized over its allowed settings
for each state, whereas $\mathcal M_8$ is fixed. After refinement to
$\mathcal M_{16}$, the finite-setting estimate becomes tighter again.}

{\begin{samepage}
\paragraph*{Finite-setting benchmarks.}
At the Werner point $r=0$, comparison with the exact steerability threshold
$1/2$~\cite{JonesWisemanDoherty2007} gives
$$
\begin{aligned}
\frac12
&<\widehat{\lambda}^{(16)}_{\rm det}=0.511
<\widehat{\lambda}^{(8)}_{\rm det}=0.527\\
&<\lambda^{(3)}_{\rm CJWR}
=\frac1{\sqrt3}\simeq0.577\\
&<\lambda^{(2)}_{\rm CJWR}=\lambda_{\rm CHSH}
=\frac1{\sqrt2}\simeq0.707.
\end{aligned}
$$
\end{samepage}
Across the sampled diagonal $X$ family in Fig.~\ref{fig:sandwich},
$\widehat{\lambda}^{(16)}_{\rm det}$ lies below the CJWR-$3$ witness
threshold. On both real-coherence families in Fig.~\ref{fig:gap}, the
thresholds satisfy
$\lambda_*=\lambda_F<\widehat{\lambda}^{(16)}_{\rm det}<
\lambda^{(3)}_{\rm CJWR}<
\lambda^{(2)}_{\rm CJWR}=\lambda_{\rm CHSH}$
at every sampled point. The strict inequalities are properties of these
examples. At $s=0.9$ on the $u=v$ family, the sixteen-direction
estimate is $0.215$, the CJWR-$3$ witness threshold is $0.285$, and the
CHSH-nonlocal threshold is $0.377$.}

\section{Relative-phase monotonicity and real-coherence bounds}
\label{sec:alignment}

The relative phase $\theta$ is the only parameter that the extension to the
full complex $X$ noise effectively adds to the real $X$ noise family. In this
section, we use the explicit boundary conditions derived above to prove
relative-phase monotonicity of the entanglement, teleportation usefulness,
CJWR witness, and CHSH-nonlocal thresholds. We then obtain lower and upper
bounds from the two real-coherence endpoints. Finally, a relative-phase scan shows how the thresholds vary with
the phase for one fixed noise family, while operational phase diagrams show
how the separable, teleportation-useless, CJWR-$3$-satisfying, and CHSH-local
intervals vary across noise states whose entanglement lies in different
blocks.

\begin{theorem}[Relative-phase monotonicity]
\label{thm:mono}
Fix $(a,b,c,d,|u|,v)$ and set $x=\cos\theta$. For the Bell mixture
$\rho_\lambda$ of Eq.~\eqref{eq:mixing-line}, the separable, teleportation-useless,
CHSH-local, and CJWR-$n$-satisfying intervals determined by the
explicit boundary conditions derived above shrink as $x$ grows.
When an interval is nonempty, its lower edge is nondecreasing and
its upper edge is nonincreasing in $x$. Consequently, the five definitive
thresholds $\lambda_*$, $\lambda_F$, $\lambda^{(3)}_{\rm CJWR}$,
$\lambda^{(2)}_{\rm CJWR}$, and $\lambda_{\rm CHSH}$ of
Eq.~\eqref{eq:definitive} are nonincreasing in $x$, equivalently
nondecreasing in $\theta\in[0,\pi]$.
\end{theorem}

Detailed proofs of these monotonicity claims are given in
Appendix~\ref{app:mono}. Here, we sketch the main steps and then discuss how to
interpret the result. Set $x=\cos\theta$ and keep the populations
$a,b,c,d$ and coherence magnitudes $|u|$ and $v=|v|$ fixed. Of the three
quantities $Q$, $P$, and $t_z$ that determine the correlation-tensor singular
values, only $Q(\lambda,x)$ depends on the relative phase. We therefore examine
its square:
$$
Q^2(\lambda,x)=\lambda^2+4\lambda(1-\lambda)|u|x+4(1-\lambda)^2|u|^2.
$$
Assume $-1\le x_1\le x_2\le1$. Then,
$$
Q^2(\lambda,x_2)-Q^2(\lambda,x_1)=4\lambda(1-\lambda)|u|(x_2-x_1)\ge0 .
$$
Because $Q\ge0$, the monotonicity of $Q^2$ in $x$ implies the
monotonicity of $Q$ in $x$. Moreover, $P=2(1-\lambda)v\ge0$ and neither $P$ nor
$t_z$ depends on $x$. Hence, $Q+P\ge0$, so both
$2(Q^2+P^2)$ and $(Q+P)^2+t_z^2$ are nondecreasing in $x$. These are the two
branches of $M$ in Eq.~\eqref{eq:chsh-max}. Then, Eq.~\eqref{eq:cjwr-F}
shows that the nonnegative CJWR values $F_3$ and $F_2$ are also nondecreasing
in $x$. If, for a Bell weight $\lambda$, $\rho_\lambda$ is CHSH-local at
$x_2$, then
$$
M(\lambda,x_1)\le M(\lambda,x_2)\le1,
$$
so $\rho_\lambda$ is also CHSH-local at $x_1$. Therefore, the CHSH-local
set at $x_2$ is contained in the CHSH-local set at $x_1$. By
Proposition~\ref{prop:windows}, these sets are intervals, so the CHSH-local
interval can only shrink as $x$ grows, equivalently widen as $\theta$ grows
from $0$ to $\pi$. Its upper edge, the definitive threshold
$\lambda_{\rm CHSH}$, is therefore nonincreasing in $x$, equivalently
nondecreasing in $\theta\in[0,\pi]$. The proofs for the CJWR-$n$-satisfying,
separable, and teleportation-useless intervals are similar and are given in
detail in Appendix~\ref{app:mono}. They give the same monotonicity for the
other four definitive thresholds.

At fixed populations and coherence magnitudes, increasing $\theta$ from $0$
to $\pi$ reduces the constructive addition of the Bell coherence and the
noise coherence in the $\Phi$ block. Consequently, a larger Bell weight can
be required to recover an ability at the definitive upper edge of its
ability-absence interval.

\begin{corollary}[Real-coherence bounds]
\label{cor:bounds}
For each operational ability in Theorem~\ref{thm:mono}, when
$a,b,c,d,|u|$, and $v$ are fixed, the corresponding definitive threshold
$\lambda(\theta)$ satisfies
$$
\lambda(0)
\;\le\;\lambda(\theta)
\;\le\;\lambda(\pi).
$$
\end{corollary}

At fixed $a,b,c,d,|u|$, and $v$, the real-coherence endpoints $u=+|u|$
and $u=-|u|$ give, respectively, the lower and upper bounds on each definitive
threshold. The bounds may be coarse, but the real-coherence formulas express
them directly, without solving a quartic equation.

\begin{figure}[!t]
\centering
\includegraphics[width=\columnwidth]{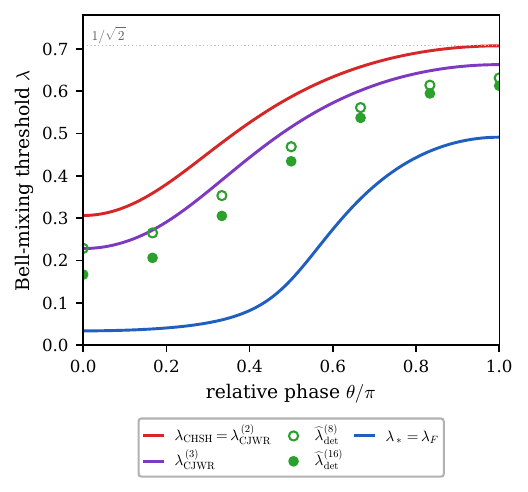}
\caption{Relative-phase scan at fixed populations and coherence
magnitudes, obtained from the $u=v$ family of Fig.~\ref{fig:gap}(b) at
coherence fraction $s=0.95$. The diagonal $(0.30,0.25,0.25,0.20)$ and
$|u|=v\approx0.233$ are fixed, while the relative phase in
$u=|u|e^{i\theta}$ varies. The definitive thresholds determined by the
explicit boundary conditions increase monotonically with
$\theta$. Here
$\lambda_*=\lambda_F$ because $b=c$~\cite{eac-paper}. Green markers show the numerical estimates
$\widehat{\lambda}^{(8)}_{\rm det}$ and
$\widehat{\lambda}^{(16)}_{\rm det}$.
At $\theta=\pi$, the CHSH-nonlocal threshold returns to the Werner value
$1/\sqrt2$.}
\label{fig:alignment}
\end{figure}

\begin{figure*}[t]
\centering
\includegraphics[width=0.98\textwidth]{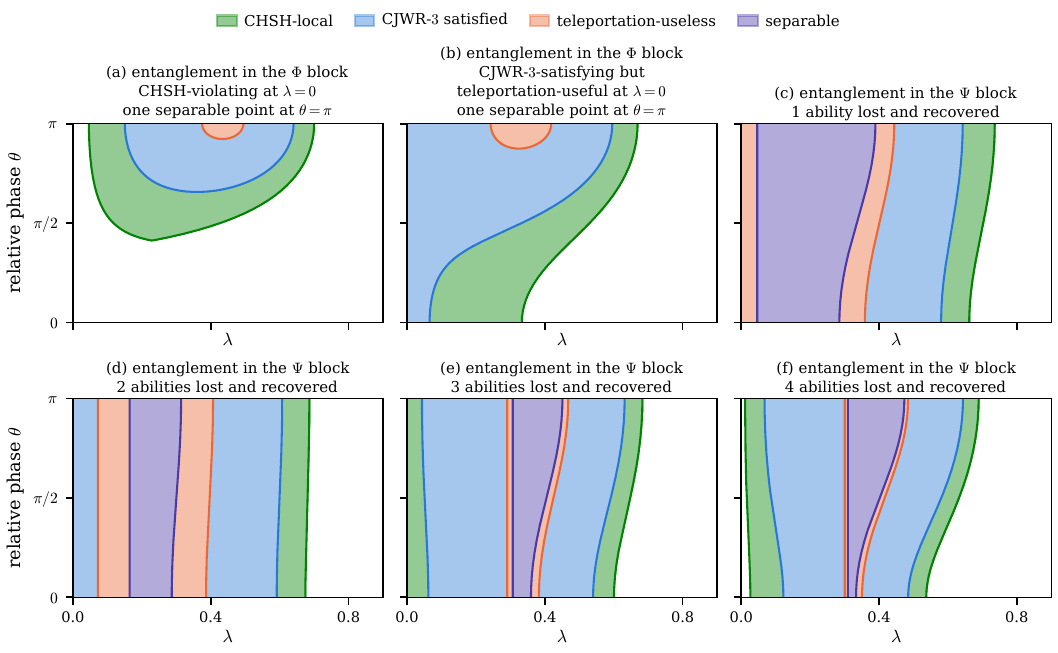}
\caption{Operational phase diagrams for Bell mixtures with $X$ noise
whose entanglement lies in either the $\Phi$ or $\Psi$ block. Colors mark
where each ability is absent, as in Fig.~\ref{fig:window-widening}. In
panels~(a) and~(b), $\sigma_X$ has entanglement in the $\Phi$ block spanned by
$|00\rangle$ and $|11\rangle$. At $\theta=0$, $\rho_\lambda$ remains
entangled for every $\lambda$ in both panels. At $\theta=\pi$, each panel has
a single Bell weight at which $\rho_\lambda$ is separable. At $\lambda=0$,
$\sigma_X$ is CHSH-violating in panel~(a), whereas it is
CJWR-$3$-satisfying but teleportation-useful in panel~(b). The CHSH-local
boundary in panel~(a) is nonsmooth where the two branches of $M$ in
Eq.~\eqref{eq:chsh-max} exchange dominance, but the CHSH-local region remains
convex. In panels~(c)--(f), $\sigma_X$
has entanglement in the $\Psi$ block spanned by $|01\rangle$ and
$|10\rangle$. At $\lambda=0$, these panels have, respectively, one, two,
three, and four operational abilities. As $\lambda$ increases, each ability
present at $\lambda=0$ is lost and later recovered. The boundary curves are
obtained from the explicit boundary conditions in the main text.}
\label{fig:gallery}
\end{figure*}

\paragraph*{Examples in the diagram.}
Figure~\ref{fig:gallery} uses selected noise states with different
populations $a,b,c,d$ and coherence magnitudes $|u|$ and $v$ to illustrate
several possible ability-absence interval structures. Panels~(a) and~(b) use
noise states with entanglement in the $\Phi$ block, for which $|u|^2>bc$.
Panels~(c)--(f) use noise states with entanglement in the $\Psi$ block, for
which $|v|^2>ad$. The examples display intervals that start at $\lambda=0$
or are detached from that endpoint, together with different relative
positions of the separable and teleportation-useless boundaries.

The significance of the phase-widening effect depends on the noise state
because the relative phase enters the boundary conditions only through
$Q(\lambda,x)$. In particular, the effect can be weak when $|u|$ is small.
Panels~(e) and~(f) use parameters for which the widening is visible while
retaining the three- and four-ability patterns at $\lambda=0$.

\paragraph*{Beyond the explicit boundary conditions.}
The arguments above use the explicit partial-transpose, fully entangled fraction,
optimized CJWR, and optimized CHSH boundary conditions available for the $X$
family. For boundaries for which no closed form is known for generic
full-rank mixed $X$ noise, such as the projective-measurement steerability
threshold of
Sec.~\ref{sec:true-steering}, we do not establish a theorem for their phase
dependence.
Ref.~\cite{PhaseOrderingPRL} proves the same phase widening for the
exact unsteerable and Bell-local intervals, without requiring
closed-form boundary equations.

\section{Product-noise tensor laws and exact boundary steering}
\label{sec:product-law}

Product noise $\sigma=\sigma_A\otimes\sigma_B$ describes independent
disturbances of the two qubits. A general product state need not have $X$
form. Its correlation tensor has rank at most one.
Proposition~\ref{prop:product} shows that the Bell mixture therefore has the
singular-value flow of an effective complex-$X$ line. We then derive the
dependence of the CHSH-nonlocal, CJWR-$3$ witness, and
{teleportation usefulness} thresholds for the corresponding Bell mixture on
the local Bloch vectors (Appendix~\ref{app:product}).
Proposition~\ref{prop:product-pure-factor} proves that, if at least one local
noise state is pure, $C(\rho_\lambda)=\lambda$ and the steerability thresholds in both
directions are zero for projective measurements and POVMs.

\begin{proposition}[Effective $X$ singular-value flow]
\label{prop:product}
For product noise $\sigma=\sigma_A\otimes\sigma_B$ with local Bloch vectors
$\vec r_A=r_A\vec n_A$ and $\vec r_B=r_B\vec n_B$, where
$r_A,r_B\in[0,1]$ and $\vec n_A,\vec n_B$ are unit vectors, the
{correlation tensor is
$T_\sigma=\vec r_A\vec r_B^{\mathsf T}
=r_Ar_B\,\vec n_A\vec n_B^{\mathsf T}$ and has rank at most one.}
Write $g=r_Ar_B$, choose
$\varphi\in[0,\pi]$, and define
$\cos\varphi=\vec n_A\cdot D\vec n_B$, where $D=\mathrm{diag}(1,-1,1)$ is
the correlation tensor of $\PhiP$. When $g=0$, the unit direction of a zero Bloch vector and
$\varphi$ may be chosen arbitrarily because the formulas below are independent
of them. The singular values of
$T(\rho_\lambda)=\lambda D+(1-\lambda)T_\sigma$ then have the closed form
(Appendix~\ref{app:product})
\begin{equation}
s_\pm(\lambda)=\bigl|\widetilde Q\pm\widetilde P\bigr|,
\qquad
s_0(\lambda)=\lambda,
\label{eq:product-flow}
\end{equation}
with
$\widetilde Q=|\lambda+(1-\lambda)(g/2)e^{i\varphi}|$ and
$\widetilde P=(1-\lambda)g/2$. This matches the complex-$X$ flow
{in Eq.~\eqref{eq:sv-flow}} with effective parameters
\begin{equation}
u_{\rm eff}=\tfrac{g}{4}\,e^{i\varphi},
\qquad
v_{\rm eff}=\tfrac{g}{4},
\qquad
\gamma_{\rm eff}=0 .
\label{eq:product-dictionary}
\end{equation}
\end{proposition}

\paragraph*{Dependence on $g$ and $\varphi$.}
The singular values of $T(\rho_\lambda)$ depend on the product noise only
through $(g,\varphi)$. As in Sec.~\ref{sec:alignment}, the optimized CHSH and
CJWR values and the fully entangled fraction are nondecreasing in
$\cos\varphi$. The corresponding thresholds are therefore nonincreasing in
$\cos\varphi$. The aligned orientation $\vec n_B=D\vec n_A$ and the
anti-aligned orientation $\vec n_B=-D\vec n_A$ therefore give, respectively,
the lower and upper bounds on these thresholds.

\begingroup

\begin{proposition}
\label{prop:product-pure-factor}
Let
$$
\rho_\lambda
=\lambda\PhiP
+(1-\lambda)\sigma_A\otimes\sigma_B,
\qquad 0\le\lambda\le1,
$$
where $\sigma_A$ and $\sigma_B$ are arbitrary qubit states. If at least
one of $\sigma_A$ and $\sigma_B$ is pure, then
\begin{equation}
C(\rho_\lambda)=\lambda.
\label{eq:product-pure-factor-concurrence}
\end{equation}
Every $\lambda>0$ gives a state that is steerable in both directions using
projective measurements. Consequently,
\begin{equation}
\resizebox{0.86\columnwidth}{!}{$\displaystyle
\lambda_*=\lambda_{\rm steer}^{A\to B}=\lambda_{\rm steer}^{B\to A}
=\lambda_{\rm steer,POVM}^{A\to B}=\lambda_{\rm steer,POVM}^{B\to A}=0.$}
\label{eq:product-pure-factor-thresholds}
\end{equation}
\end{proposition}

The concurrence identity follows from Wootters' formula. The steerability
statement follows from Lemma~\ref{lem:product-null} and
Theorem~\ref{cor:pure-noise-steering}.
Details are given in Appendix~\ref{app:product}.
\endgroup

\paragraph*{Scope of the product-noise result.}
Proposition~\ref{prop:product-pure-factor} determines the steerability
thresholds for projective measurements and POVMs when at least one of the
local noise states $\sigma_A$ and $\sigma_B$ is pure.
When both local noise states are full rank, the correlation tensor
determines the optimized CHSH and CJWR witness thresholds, but it does not necessarily
determine the exact directional projective-measurement steerability threshold.
Ref.~\cite{SteeringMarginalsLetter} gives two product-noise mixtures with the
same correlation tensor but different $A\to B$ steerability thresholds.

\section{Applications: operational loss under local channels and entanglement swapping}
\label{sec:death-times}

Along any trajectory that keeps the evolving state within the $X$ family,
the correlation tensor gives closed-form expressions for its singular values,
the CHSH and CJWR witness values, and the fully entangled fraction. Concurrence is also available
in closed form for $X$ states. We use these formulas to track CHSH
nonlocality, optimized three-setting CJWR violation, standard teleportation
usefulness, and entanglement as functions of time under local noise and of
the number of swaps under repeated post-selected entanglement swapping.

\subsection{Local-channel evolution}

\begin{figure*}[!t]
\centering
\includegraphics[width=0.92\textwidth]{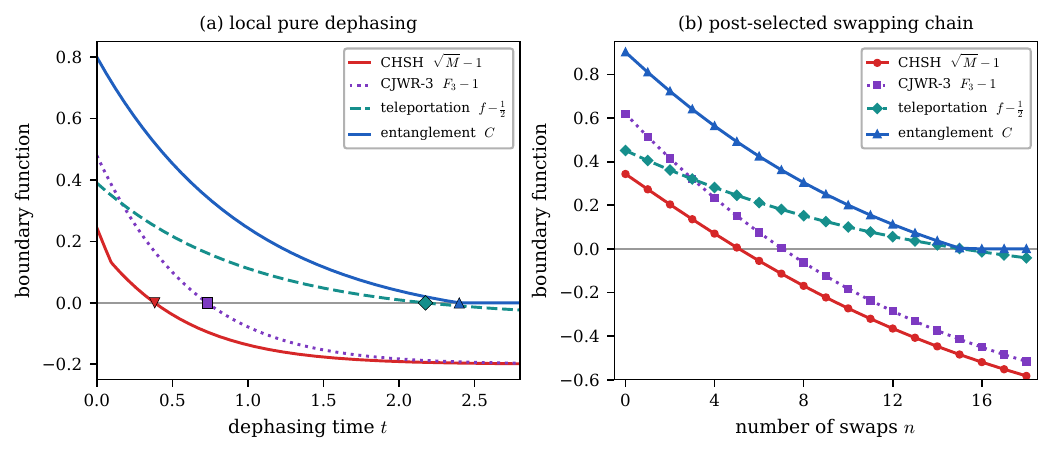}
\caption{Boundary functions under (a) local pure dephasing and
(b) a post-selected entanglement-swapping chain. Only the loss points are
compared because the four functions have different normalizations. In panel (a),
optimized CHSH violation, optimized three-setting CJWR violation, standard
teleportation usefulness, and entanglement are lost at the respective times
$t\simeq0.383$, $0.730$, $2.175$, and $2.398$. In panel~(b), their
ability-loss depths are $n_{\rm CHSH}=6$,
$n_{\mathrm{CJWR}\text{-}3}=8$, and $n_F=n_*=16$. Standard
teleportation usefulness and entanglement are lost at the same
swap depth for this symmetric real-$X$ sequence. Connecting lines
are visual guides, and panel (b) does not include the cumulative postselection
probability.}
\label{fig:death}
\end{figure*}

Let $\mathcal E_A(t)$ and $\mathcal E_B(t)$ be the single-qubit noise
channels describing the noise accumulated on $A$ and $B$ up to time $t$, with
$\mathcal E_A(0)=\mathcal E_B(0)=\mathrm{id}$. For an initial $X$ state
$\rho(0)$, let
\begin{equation}
\rho(t)=\bigl[\mathcal E_A(t)\otimes\mathcal E_B(t)\bigr][\rho(0)].
\label{eq:full-channel-trajectory}
\end{equation}
We consider channels for which $\rho(t)$ remains in the $X$ family for
every $t\ge0$. Examples are given in Appendix~\ref{app:channels}. If
$\rho(0)$ has one of the four abilities above, denoted by $\mathcal O$, define
its ability-loss time by
\begin{equation}
t_{\mathcal O}
:=\inf\{t\ge0:
\rho(t)\text{ lacks }\mathcal O\},
\quad \inf\emptyset:=\infty,
\label{eq:death-time}
\end{equation}
To locate the loss point of each ability, we use the following four
functions:
\begin{equation}
\begin{aligned}
\Delta_{\rm CHSH}(\rho)&=\sqrt{M(\rho)}-1,
\\
\Delta_{\mathrm{CJWR}\text{-}3}(\rho)&=F_3(\rho)-1,\\
\Delta_F(\rho)&=f(\rho)-\tfrac12,\\
\Delta_*(\rho)&=C(\rho).
\end{aligned}
\label{eq:loss-functions}
\end{equation}
In the order displayed, the four functions are positive exactly when $\rho$ is
CHSH-nonlocal, violates the
optimized three-setting CJWR inequality, has standard teleportation
usefulness, and is entangled. Therefore, the first zero along the
continuous local-noise trajectory gives the corresponding ability-loss time. Because
the four functions have
different normalizations, Fig.~\ref{fig:death} compares these loss points, not their
magnitudes.

Figure~\ref{fig:death}(a) applies the local pure-dephasing channel
$\mathcal D_{\gamma_\varphi(t)}$ to both $A$ and $B$ independently, where
$\gamma_\varphi(t)=e^{-t}$. The initial point is
$(a,b,c,d,u,v)=(0.45,0.08,0.02,0.45,0.44,0)$. Along the trajectory, the only
change is $u\mapsto ue^{-t}$. The loss points are
$t_{\rm CHSH}\simeq0.383$,
$t_{\mathrm{CJWR}\text{-}3}\simeq0.730$,
$t_F\simeq2.175$, and
$t_*\simeq2.398$. Thus, optimized CHSH violation, optimized three-setting
CJWR violation, standard teleportation usefulness, and entanglement disappear
successively. Other $X$-preserving channels are treated by substituting their
evolved matrix entries into the same formulas. This comparison extends
channel studies of entanglement sudden death and the successive loss of CHSH,
CJWR, and entanglement~\cite{YuEberly2004,AlmeidaEtAl2007Science,
CostaBeimsAngelo2016} by including standard teleportation usefulness.

\begin{figure}[!ht]
\centering
\includegraphics[width=0.96\columnwidth]{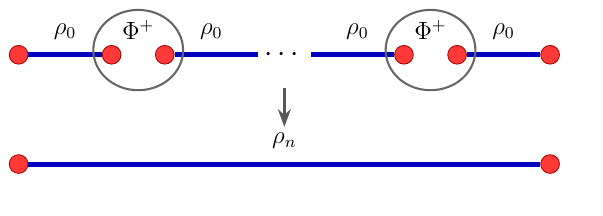}
\caption{A post-selected chain with $n$ swaps. The upper row shows
$n+1$ identical elementary links and the retained $\PhiP$ outcome
at each of the $n$ Bell measurements. The lower row shows the end-to-end
output after $n$ swaps.}
\label{fig:swapping-chain}
\end{figure}

\subsection{Post-selected swapping chain}

Figure~\ref{fig:swapping-chain} illustrates $n$ swaps across $n+1$
elementary links, each described by the same two-qubit state $\rho_0$. The
protocol aims to establish end-to-end entanglement between the two end nodes
by post-selected entanglement swapping across these
links~\cite{ZukowskiEtAl1993}. At each step, a Bell-basis measurement is
performed on the two middle qubits of two adjacent links, and only the
outcome $\PhiP$ is retained. The measured qubits are discarded, and
the state of the two outer qubits is the output of the process.

Applying one swapping step to two copies of $\rho_0$
gives the endpoint state $\rho_1$. More generally, $\rho_n$ denotes the
endpoint state after $n$ swaps across $n+1$ elementary links, with the
outcome $\PhiP$ retained at every Bell measurement. A recurrence for
the sequence $\{\rho_n\}$ can be worked out directly. If the two inputs to a
swapping step have $X$ form, the output
also has $X$ form~\cite{MunozGruningRoa2014}.

For Figure~\ref{fig:death}(b), we set $\gamma_1=0.95$ and apply the
amplitude-damping channel $\mathcal A_{\gamma_1}$ independently to both
qubits of $\PhiP$, with no pure dephasing. The elementary-link state
is
\begin{equation}
\rho_0
=\bigl(\mathcal A_{\gamma_1}\otimes\mathcal A_{\gamma_1}\bigr)
  (\PhiP).
\label{eq:swap-link}
\end{equation}
If $\rho_0$ has the ability $\mathcal O$, define its ability-loss depth by
$$
n_{\mathcal O}
:=\min\{n\in\mathbb Z_{\ge0}:\rho_n\text{ lacks }\mathcal O\},
\qquad \min\emptyset:=\infty.
$$
Direct evaluation gives
$$
n_{\rm CHSH}=6,\qquad
n_{\mathrm{CJWR}\text{-}3}=8,\qquad
n_F=n_*=16.
$$
The cumulative postselection success probability is exponentially small
in $n$ for this sequence. We present detailed
calculations in Appendix~\ref{app:swapping}, where we also show why standard
teleportation usefulness and entanglement are lost together at the same swap
depth in this specific setup.

\begin{table*}[t]
\caption{Operational thresholds along the Bell-mixing line
$\rho_\lambda=\lambda\PhiP+(1-\lambda)\sigma_X$.
All entries use the definitive-threshold convention of
Eq.~\eqref{eq:definitive}. Whenever a threshold is determined, the
corresponding ability-absence interval is also determined. Here $C$ is the
concurrence, $f$ is the fully
entangled fraction, $M$ is the optimized CHSH quantity, and $F_n$ is the
optimized $n$-setting CJWR value. The final column lists data on $\sigma_X$
that suffice to determine the threshold. The four correlators
$(T_{\sigma_X})_{ij}=\langle\sigma_i\otimes\sigma_j\rangle_{\sigma_X}$ with
$i,j\in\{x,y\}$, together with
$(T_{\sigma_X})_{zz}=\langle\sigma_z\otimes\sigma_z\rangle_{\sigma_X}$,
determine its correlation tensor. When $\sigma_X$ is real,
$T_{\sigma_X}$ is diagonal. Together, these five correlators and the two
local expectation values $\langle\sigma_z\otimes I\rangle_{\sigma_X}$ and
$\langle I\otimes\sigma_z\rangle_{\sigma_X}$ determine all entries of
$\sigma_X$.}
\label{tab:thresholds}
\begingroup
\setlength{\tabcolsep}{2.5pt}
\renewcommand{\arraystretch}{1.35}
\newcommand{\TCRR}{\raggedright\rightskip=0pt plus 2em\relax
  \parfillskip=0pt plus 1fil\relax
  \hyphenpenalty=50\relax\emergencystretch=1em\relax}
\newcommand{\TCPT}{\vspace{2.5pt}}
\newcommand{\TCPB}{\par\vspace{2.5pt}}
\newcommand{\TCA}[1]{\parbox[c]{0.115\textwidth}{\TCPT\centering #1\TCPB}}
\newcommand{\TCB}[1]{\parbox[c]{0.130\textwidth}{\TCPT\centering #1\TCPB}}
\newcommand{\TCC}[1]{\parbox[c]{0.115\textwidth}{\TCPT\centering #1\TCPB}}
\newcommand{\TCD}[1]{\parbox[c]{0.370\textwidth}{\TCPT\TCRR #1\TCPB}}
\newcommand{\TCE}[1]{\parbox[c]{0.200\textwidth}{\TCPT\TCRR #1\TCPB}}
\begin{tabular}{|l|l|l|l|l|}
\hline
\TCA{Threshold type}
& \TCB{Threshold condition}
& \TCC{Threshold}
& \TCD{\centering Threshold formula or status}
& \TCE{\centering Sufficient data}\\
\hline
\TCA{Entanglement}
& \TCB{PPT (equivalently, $C=0$)}
& \TCC{$\lambda_*$}
& \TCD{Proposition~\ref{prop:windows} gives the entanglement threshold in
closed form for all $X$ noise. Ref.~\cite{eac-paper} gives the threshold for
arbitrary product noise.}
& \TCE{Full information about $\sigma_X$}\\
\hline
\TCA{Teleportation usefulness}
& \TCB{$f\le\tfrac12$}
& \TCC{$\lambda_F$}
& \TCD{Proposition~\ref{prop:windows} gives the teleportation usefulness
threshold in closed form for all $X$ noise. For product noise,
Appendix~\ref{app:product} derives the threshold from the effective $X$-state
singular-value flow and recovers the results of
Ref.~\cite{eac-paper}.}
& \TCE{Full correlation tensor of $\sigma_X$}\\
\hline
\TCA{Steerability}
& \TCB{Local-hidden-state model for the stated direction and measurement
class}
& \TCC{$\begin{gathered}
\lambda_{\rm steer}^{A\to B}\\[2pt]
\lambda_{\rm steer,POVM}^{A\to B}\\[2pt]
\lambda_{\rm steer}^{B\to A}\\[2pt]
\lambda_{\rm steer,POVM}^{B\to A}
\end{gathered}$}
& \TCD{All four thresholds equal $\lambda_*$ for arbitrary pure noise,
product noise with a pure local factor, and the two diagonal mixed $X$
families (Theorem~\ref{cor:pure-noise-steering},
Propositions~\ref{prop:boundary-steering} and~\ref{prop:product-pure-factor},
and Corollary~\ref{cor:povm-projective}). For product noise with a pure local
factor and for the two diagonal mixed $X$ families, the common threshold is
zero. For generic mixed $X$ noise, no exact formula is known, and finite-setting
semidefinite programs give upper bounds on the projective-measurement
steerability thresholds.}
& \TCE{Full information about $\sigma_X$. The correlation tensor alone does
not determine the directional thresholds~\cite{SteeringMarginalsLetter}.}\\
\hline
\TCA{CJWR-$3$ witness}
& \TCB{$F_3\le1$}
& \TCC{$\lambda^{(3)}_{\rm CJWR}$}
& \TCD{For all $X$ noise, Theorem~\ref{thm:cjwr} and
Appendix~\ref{app:chsh} determine the CJWR-$3$ witness threshold from a
quadratic boundary equation. For product noise,
Appendix~\ref{app:product} derives the threshold from the effective $X$-state
singular-value flow.}
& \TCE{Full correlation tensor of $\sigma_X$}\\
\hline
\TCA{CHSH nonlocality and CJWR-$2$ witness}
& \TCB{$M\le1$, equivalently $F_2\le1$}
& \TCC{$\lambda_{\rm CHSH}=\lambda^{(2)}_{\rm CJWR}$}
& \TCD{For all $X$ noise, Theorem~\ref{thm:chsh} and
Appendix~\ref{app:chsh} determine the common CHSH-nonlocal and CJWR-$2$
witness threshold from explicit boundary equations.
For product noise, Appendix~\ref{app:product} derives the common threshold
from the effective $X$-state singular-value flow.}
& \TCE{Full correlation tensor of $\sigma_X$}\\
\hline
\end{tabular}
\endgroup
\end{table*}

\section{Discussion and conclusion}
\label{sec:conclusion}

Table~\ref{tab:thresholds} summarizes the results discussed
above, the available formulas or conditions, and the data sufficient to
determine them.

\paragraph*{{Ability-absence intervals.}}
{{Each operational ability considered in this paper is absent on a closed
convex set of states. Under the affine map $\lambda\mapsto\rho_\lambda$, its
preimage is empty, a single point, or a closed interval with two distinct
endpoints. If a nondegenerate interval begins at $\lambda=0$, its upper edge is
the single onset of the ability. If it is detached from $\lambda=0$, its lower
and upper edges mark the loss and recovery of the ability, respectively.}
Positivity of the partial
transpose or, equivalently, vanishing concurrence determines the separable
interval. {The correlation-tensor singular-value flow gives the fully
entangled fraction and hence the teleportation-useless interval. It also
determines the CHSH-local and CJWR-satisfying intervals.} The explicit boundary
conditions establish relative-phase widening of all four intervals.
Using the fact that the correlation tensor of a product state has rank at
most one, we obtain the product-noise teleportation usefulness, CJWR, and CHSH
thresholds from the corresponding thresholds of an effective complex-$X$
noise state.}

\paragraph*{{Exact steerability and scope.}}
{The projective-measurement steerability threshold generally requires
information beyond the correlation tensor. Section~\ref{sec:true-steering}
determines it exactly for arbitrary pure two-qubit noise and for two diagonal
mixed-$X$ noise families. Proposition~\ref{prop:product-pure-factor} further
shows that the projective-measurement and POVM steerability thresholds are zero
in both directions for product noise with at least one pure local noise state.
For generic full-rank mixed $X$ noise, {the finite-setting semidefinite
programs give upper bounds on the exact projective-measurement steerability
threshold.} Corollary~\ref{cor:order} places the
{definitive projective-measurement steerability threshold below the
definitive CJWR witness thresholds. The definitive thresholds for standard
teleportation usefulness and projective-measurement steerability are not
mutually ordered.}
{Ref.~\cite{PhaseOrderingPRL} proves that the exact steerability threshold
has the same relative-phase monotonicity established in
Sec.~\ref{sec:alignment} for the explicit thresholds.}}

\paragraph*{{Relation to previous results.}}
{{Ref.~\cite{eac-paper} treats the entanglement and teleportation
usefulness thresholds for separable $X$ noise and arbitrary product noise. For
the $X$-noise sector, our complete ability-absence intervals extend the
analysis to arbitrary $X$ noise, whether separable or entangled. For separable
$X$ noise, the definitive upper edges of the intervals recover both thresholds of
Ref.~\cite{eac-paper}. For arbitrary product noise, our fully entangled-fraction calculation
independently recovers the teleportation usefulness threshold, while the
correlation-tensor reduction adds the optimized CJWR and CHSH witness thresholds.
{Together with the product-noise entanglement threshold of
Ref.~\cite{eac-paper}, which is not recovered here, these results determine
the entanglement, teleportation usefulness, CJWR-$3$ witness, and
CHSH-nonlocal definitive thresholds for arbitrary product noise.}}}

\paragraph*{{Information contained in the correlation tensor.}}
For an arbitrary $X$ state $\rho_\lambda$, the four $xy$-block entries
$T_{ij}(\lambda)$ with $i,j\in\{x,y\}$, together with $t_z(\lambda)$,
{determine the
correlation tensor and therefore its three singular values and determinant.
These quantities give $M$, $F_2$, $F_3$, and $f$. Thus, at fixed $\lambda$, at
most five two-body Pauli correlators determine the optimized CHSH and CJWR
values and the fully entangled fraction.} {Along the full Bell-mixing line,
the closed-form singular-value flow in $\lambda$ determines the teleportation
usefulness, CHSH-nonlocal, and CJWR witness thresholds.}

{{Ref.~\cite{SteeringMarginalsLetter} gives two complementary examples. In
the full-rank product-noise example, exchanging the two local Bloch radii preserves the
correlation tensor but changes the $A\to B$ projective-measurement steerability
threshold by more than $1/100$. {In the full-rank entangled $X$-noise
example, the two states have the same correlation tensor, spectrum, and
concurrence, while their local expectation values
$\langle\sigma_z\otimes I\rangle$ and $\langle I\otimes\sigma_z\rangle$ are
exchanged. Their definitive thresholds nevertheless differ by more
than $1/12$.} {Beyond the discussion on the correlation tensor,
Ref.~\cite{PhaseOrderingPRL} proves that, among all noise states with a fixed
$X$ part $\sigma_X$, the state $\sigma_X$ is the unique $X$ state and has the
largest definitive threshold for every operational ability considered in our
paper.}}}

{{Further progress in characterizing exact directional steerability thresholds
will provide deeper insight into quantum correlations, whose structure remains
rich even in two-qubit systems.}}

\begin{acknowledgments}
The author specified the semidefinite program, solver settings, and
validation tests and used OpenAI Codex with GPT-5.6 (accessed August 2026) to assist with
implementing, testing, and debugging. The author checked
all reported numerical values against the validation scripts and saved solver
outputs. The author also specified the intended content of each figure, used
Codex to assist with LaTeX syntax and detailed figure formatting, and
inspected the final figures.
\end{acknowledgments}

\section*{Data Availability}
The numerical data and custom scripts supporting the findings of this
article are not publicly archived because the associated research program is
ongoing. They are available from the author upon reasonable request.

\appendix

\section{Gauge reduction and the singular-value flow}
\label{app:gauge}

We show how local phase rotations act on the coherences, evaluate the
correlation tensor along the complex $X$ Bell-mixing line, and prove the
singular-value flow of Proposition~\ref{prop:flow}.

\paragraph*{{Local phase rotations.}}
In the computational basis, the local phase unitary
$W=\mathrm{diag}(1,e^{i\phi_A})\otimes\mathrm{diag}(1,e^{i\phi_B})$
has diagonal entries
$(1,e^{i\phi_B},e^{i\phi_A},e^{i(\phi_A+\phi_B)})$.
Conjugation by $W$, $\rho\mapsto W\rho W^\dagger$, transforms the
coherences $u$ and $v$ in Eq.~\eqref{eq:x-matrix} as
$$
u\mapsto e^{-i(\phi_A+\phi_B)}u,
\qquad
v\mapsto e^{-i(\phi_A-\phi_B)}v.
$$
The Bell coherence
$\langle00|\PhiP|11\rangle$ acquires the same factor
$e^{-i(\phi_A+\phi_B)}$ as $u$. Choosing
$\phi_B=-\phi_A=-\phi$ fixes $\PhiP$ and $u$ while
rotating $v$ by $e^{-2i\phi}$, which proves that {$\arg v$ can be removed
from the entire Bell-mixing line by a local unitary}. Choosing
$\phi_B=\phi_A$ instead multiplies $u$ and the Bell coherence by the
same factor. Their common phase shifts, but their phase difference
stays fixed. This difference is the physical relative phase, and in the gauge
where the Bell coherence is real and positive it equals $\theta=\arg u$.

\paragraph*{Correlation tensor.}
The correlation tensor $T_{ij}=\mathrm{Tr}[\rho\,(\sigma_i\otimes\sigma_j)]$
is linear in $\rho$. Therefore, along the mixing line in
Eq.~\eqref{eq:mixing-line},
$T(\lambda)=(1-\lambda)\,T_{\sigma_X}+\lambda\,T_{\PhiP}$.
Evaluating the endpoints suffices. Write $u=u_R+i u_I$ and
$v=v_R+i v_I$, where $u_R,u_I,v_R,v_I\in\mathbb R$.
With $\gamma:=a-b-c+d$, direct calculation gives the correlation tensor
of $\sigma_X$,
\begin{equation}
T_{\sigma_X}=
\begin{pmatrix}
2(u_R+v_R) & 2(v_I-u_I) & 0\\
-2(u_I+v_I) & 2(v_R-u_R) & 0\\
0 & 0 & \gamma
\end{pmatrix}.
\label{eq:complex-x-tensor}
\end{equation}
The Bell state $\PhiP$ is also an $X$ state, and
$T_{\PhiP}=\mathrm{diag}(1,-1,1)$. Let $J(\lambda)$ denote the upper-left
$2\times2$ block of $T(\lambda)$. The $z$ component of $T(\lambda)$ is
$t_z(\lambda)=(1-\lambda)\gamma+\lambda=\gamma+\lambda(1-\gamma)$.
The latter identity is Eq.~\eqref{eq:tz}.

\paragraph*{Singular values.}
The $z$ component $t_z(\lambda)$ of $T(\lambda)$ directly gives the
singular value $|t_z(\lambda)|$. The remaining two singular values are those
of $J(\lambda)$. These singular values are the nonnegative
square roots of the eigenvalues of $J^{\mathsf T}J$. For a $2\times2$ matrix, these two
eigenvalues sum to $\mathrm{Tr}(J^{\mathsf T}J)$ and multiply to
$\det(J^{\mathsf T}J)=(\det J)^2$. Hence, the singular values
$s_1,s_2\ge0$ obey
$s_1^2+s_2^2=\mathrm{Tr}(J^{\mathsf T}J)
=J_{11}^2+J_{12}^2+J_{21}^2+J_{22}^2$ and
$s_1s_2=|\det J|=|J_{11}J_{22}-J_{12}J_{21}|$. Therefore,
$(s_1+s_2)^2$ and $(s_1-s_2)^2$ are the two values
$$
\begin{aligned}
\mathrm{Tr}(J^{\mathsf T}J)+2\det J
&=(J_{11}+J_{22})^2+(J_{12}-J_{21})^2\\
&=16(1-\lambda)^2|v|^2,\\
\mathrm{Tr}(J^{\mathsf T}J)-2\det J
&=(J_{11}-J_{22})^2+(J_{12}+J_{21})^2\\
&=4\,|\lambda+2(1-\lambda)u|^2.
\end{aligned}
$$
If $\det J\ge0$, the first displayed value is $(s_1+s_2)^2$ and the
second is $(s_1-s_2)^2$. If $\det J<0$, these assignments are reversed. In
either case,
\begin{equation}
s_{1,2}=\tfrac12\Bigl|
\sqrt{\mathrm{Tr}(J^{\mathsf T}J)+2\det J}
\pm\sqrt{\mathrm{Tr}(J^{\mathsf T}J)-2\det J}\Bigr|.
\label{eq:2x2-sv}
\end{equation}
In the gauge $v\ge0$, the two square roots are
$4(1-\lambda)|v|=2P(\lambda)$ and
$2|\lambda+2(1-\lambda)u|=2Q(\lambda)$.
{Eq.~\eqref{eq:2x2-sv}} then gives $s_{1,2}=\tfrac12|2P\pm2Q|$,
{that is,} $s_\pm(\lambda)=|Q(\lambda)\pm P(\lambda)|$, which, together
with $|t_z(\lambda)|$, proves the singular-value flow. Setting
$u_I=0$ and using $v\ge0$
recovers the real-coherence statement $\{s_+,s_-\}=\{|t_x|,|t_y|\}$ of
Sec.~\ref{sec:xsetup}, with $t_x=q+P$ and $t_y=q-P$ for
$q=\lambda+2(1-\lambda)u$.

\section{Proof of the interval proposition and exact separable and
teleportation-useless intervals}
\label{app:lower}
\label{app:windows}

In this appendix, we prove the interval and nesting statements of
Proposition~\ref{prop:windows} and derive the separable and
teleportation-useless intervals for every physical complex $X$ noise state,
including entangled noise.

\paragraph*{Convexity and nesting.}
{The separable states form a closed convex set. The fully entangled fraction
$f(\rho)=\max_\Phi\langle\Phi|\rho|\Phi\rangle$, where the maximum is over
all maximally entangled states $|\Phi\rangle$, is convex~\cite{eac-paper}. It
is also continuous. For any two states $\rho$ and $\tau$, choose a maximally
entangled state $|\Phi_\rho\rangle$ that attains $f(\rho)$. Since
$f(\tau)\ge\langle\Phi_\rho|\tau|\Phi_\rho\rangle$,
$$
f(\rho)-f(\tau)
\le\langle\Phi_\rho|(\rho-\tau)|\Phi_\rho\rangle
\le\|\rho-\tau\|_\infty,
$$
where $\|\cdot\|_\infty$ is the operator norm. Exchanging $\rho$ and $\tau$
gives $|f(\rho)-f(\tau)|\le\|\rho-\tau\|_\infty$, which proves continuity.
Convexity of $f$ makes the sublevel set $\{f\le1/2\}$ convex, and continuity
makes it closed. If
$\rho_n$ is a sequence in this set with $\rho_n\to\rho$, continuity gives
$f(\rho)=\lim_{n\to\infty}f(\rho_n)\le1/2$, so $\rho$ also belongs to the
set.} The optimized CHSH and CJWR witness values maximize absolute
expectation values over their allowed measurement
settings~\cite{CHSH1969,CavalcantiJonesWisemanReid2009}. Each signed expectation
is linear in the state, so taking its absolute value and then maximizing
preserves convexity. For either optimized value $G$, the maximizing-setting
argument also gives $|G(\rho)-G(\tau)|\le C\|\rho-\tau\|_1$, where $C$ bounds
the witness-operator norms, and hence proves continuity. Their nonviolation
sets are therefore closed and convex.
{The intersection of any of these closed convex sets with the segment
$\{\rho_\lambda:0\le\lambda\le1\}$ is a single closed interval, possibly
empty.}

{The intervals are nested because every separable state is
teleportation-useless, and every teleportation-useless two-qubit state
satisfies the three-setting CJWR inequality. The second implication is the
contrapositive of the result in Ref.~\cite{FanJiaQiu2022}. For every state,
$F_3\ge F_2$ and $F_2^2=M$, where $M$ is the optimized CHSH quantity. These
relations give the nesting in Proposition~\ref{prop:windows}. At $\lambda=1$,
the state $\PhiP$ is entangled, has teleportation
usefulness, violates the CJWR-$n$ inequality, and is CHSH-nonlocal. Therefore,
none of the intervals contains $\lambda=1$.}

\paragraph*{{Bell mixture in the $\eta$ variable.}}
For $\eta=\lambda/(1-\lambda)$, the state with $\lambda<1$ can be written as
\begin{equation}
\rho_\lambda
=\frac{\sigma_X+\eta\PhiP}{1+\eta}.
\label{eq:eta-line}
\end{equation}
{The factor $1/(1+\eta)$ is positive and therefore does not affect
positivity of the partial transpose.} The numerator in
{Eq.~\eqref{eq:eta-line}} has
populations $(a+\eta/2,b,c,d+\eta/2)$ and coherences
$(u+\eta/2,v)$.

\paragraph*{Separability.}
For two qubits, separability is equivalent to positivity under partial
transpose. {Applying the two PPT conditions of
Eq.~\eqref{eq:x-separability-conditions} to the state in
Eq.~\eqref{eq:eta-line} gives}
\begin{equation}
|u+\eta/2|^2\le bc,
\qquad
|v|^2\le(a+\eta/2)(d+\eta/2).
\label{eq:eta-ppt}
\end{equation}
{The first inequality is equivalent to
$(\eta+2\,\mathrm{Re}\,u)^2\le4D_{\rm sep}$, so it can hold for real $\eta$
only when $D_{\rm sep}=bc-(\mathrm{Im}\,u)^2\ge0$. In that case, its solution
is $\eta\in[\eta_-,\eta_+]$, where $\eta_\pm$ are defined in
Eq.~\eqref{eq:sep-window}. The second inequality is equivalent to
\begin{equation}
(\eta+a+d)^2\ge(a-d)^2+4|v|^2.
\label{eq:eta-ppt-psi}
\end{equation}
On the physical range $\eta\ge0$, Eq.~\eqref{eq:eta-ppt-psi} holds for
$\eta\ge\max\{0,\eta_\Psi\}$, where $\eta_\Psi$ is also defined in
Eq.~\eqref{eq:sep-window}. The intersection of these two $\eta$ ranges is the
separable interval stated in Proposition~\ref{prop:windows}. It is empty when
its lower endpoint exceeds $\eta_+$.}

\paragraph*{Teleportation usefulness.}
Applying the two-qubit singular-value formula for the fully entangled
fraction~\cite{HorodeckiTeleportationBell1996,HorodeckiTeleportation1999} to
the flow in Proposition~\ref{prop:flow}, whose singular values are
$|Q+P|$, $|Q-P|$, and $|t_z|$ and whose determinant is
$(P^2-Q^2)t_z$, gives
\begin{equation}
f(\rho_\lambda)=\frac14\left[1+
\max\{t_z+2Q,\,2P-t_z\}\right].
\label{eq:eta-fef}
\end{equation}
Under $\lambda=\eta/(1+\eta)$, the flow has
$Q=|\eta+2u|/(1+\eta)$, $P=2|v|/(1+\eta)$, and
$t_z=(\eta+\gamma)/(1+\eta)$, where $\gamma=a-b-c+d$. Using
$a+b+c+d=1$, Eq.~\eqref{eq:eta-fef} shows that $f\le1/2$ is equivalent to
\begin{equation}
|\eta+2u|\le b+c,
\qquad
\eta\ge2|v|-(a+d).
\label{eq:eta-teleportation}
\end{equation}
The first inequality is equivalent to
$(\eta+2\,\mathrm{Re}\,u)^2\le D_F$, so it can hold for real $\eta$ only when
$D_F=(b+c)^2-4(\mathrm{Im}\,u)^2\ge0$. In that case, its solution is
$\eta\in[\eta_{F,-},\eta_{F,+}]$, where the roots are defined in
Eq.~\eqref{eq:teleportation-window}. Intersecting this range with the second
condition in Eq.~\eqref{eq:eta-teleportation} and with $\eta\ge0$ gives the
teleportation-useless interval in Proposition~\ref{prop:windows}. It is empty
when its lower endpoint exceeds $\eta_{F,+}$.

For separable noise, applying the inverse map
$\lambda=\eta/(1+\eta)$ to the upper endpoints $\eta_+$ and $\eta_{F,+}$
recovers the entanglement and teleportation usefulness thresholds of
Ref.~\cite{eac-paper}. For entangled noise, $\rho_0=\sigma_X$ is entangled, so
the separable interval is either detached from $\lambda=0$ or empty. The same
conditions classify the teleportation-useless interval as beginning at
$\lambda=0$, detached from that endpoint, or empty.

\section{\texorpdfstring{{Local channels that preserve the $X$ family and
channel commutation}}{Local channels that preserve the X family and channel
commutation}}
\label{app:channels}

We present the two channel properties used in Sec.~\ref{sec:xsetup}: local
channels with diagonal or antidiagonal Kraus operators preserve $X$ states,
and amplitude damping commutes with the pure-dephasing channel
$\mathcal D_{\gamma_\varphi}$. We also show that this commutation fails for
pure dephasing about a transverse axis, so the second property is specific to the
$z$ axis.

\paragraph*{{$X$-family preservation.}}
Write $Z=\sigma_z$. Conjugation by $Z\otimes Z$ multiplies the entry
$\rho_{ij}$ by $+1$ when $|i\rangle$ and $|j\rangle$ lie in the same pair
$\{|00\rangle,|11\rangle\}$ or $\{|01\rangle,|10\rangle\}$, and by $-1$
otherwise. A two-qubit operator therefore has $X$ form if and only if
$(Z\otimes Z)\rho(Z\otimes Z)=\rho$. By the same count on a single qubit, a
one-qubit operator $K$ is diagonal or antidiagonal if and only if
$ZKZ=\pm K$, with the plus sign for diagonal and the minus sign for
antidiagonal. A product Kraus operator $K=K_A\otimes K_B$ built from such
operators then satisfies $(Z\otimes Z)K(Z\otimes Z)=\pm K$. Since
$(Z\otimes Z)^2=I$ and $\rho$ has $X$ form,
$$
(Z\otimes Z)K\rho K^\dagger(Z\otimes Z)
=(\pm1)^2K\rho K^\dagger
=K\rho K^\dagger .
$$
The sign squares to one, so each Kraus term again has $X$ form, and so does
their sum. Hence, a product channel $\mathcal E_A\otimes\mathcal E_B$
preserves the $X$ family whenever every Kraus operator of $\mathcal E_A$ and
$\mathcal E_B$ is diagonal or antidiagonal. The amplitude-damping Kraus
operators $A_0$ and $A_1$ are, respectively, diagonal and antidiagonal. Every
one-qubit Pauli channel also admits a Kraus representation of this form.

We use the amplitude-damping channel $\mathcal A_{\gamma_1}$ defined in
Sec.~\ref{sec:xsetup}. For $U\in\{\sigma_x,\sigma_y,\sigma_z\}$ and
$0\leq p\leq\tfrac12$, define the Pauli channel
$$
\mathcal D^U_p(\rho)=(1-p)\,\rho+p\,U\rho U.
$$
For $U=\sigma_z$ and
$p=(1-\sqrt{\gamma_\varphi})/2$, this channel is
$\mathcal D_{\gamma_\varphi}$ of the main text.

\paragraph*{{$\sigma_z$: commuting.}} On
$\rho=\bigl(\begin{smallmatrix}\rho_{00}&\rho_{01}\\
\rho_{10}&\rho_{11}\end{smallmatrix}\bigr)$, both orders return
$$
\begin{pmatrix}
\rho_{00}+(1-\gamma_1)\rho_{11}&(1-2p)\sqrt{\gamma_1}\,\rho_{01}\\
(1-2p)\sqrt{\gamma_1}\,\rho_{10}&\gamma_1\rho_{11}
\end{pmatrix}.
$$
The populations are untouched by $\mathcal D_p^{\sigma_z}$, and the two
coherence factors multiply in either order.

\paragraph*{{$\sigma_x$: not commuting.}} The bit flip swaps the levels,
$$
\sigma_x A_0\sigma_x=\mathrm{diag}(\sqrt{\gamma_1},\,1),
\qquad
\sigma_x A_1\sigma_x=\sqrt{1-\gamma_1}\,|1\rangle\langle0|,
$$
turning relaxation $|1\rangle\to|0\rangle$ into
excitation $|0\rangle\to|1\rangle$.
At $\rho=|1\rangle\langle1|$ and $\gamma_1=p=\tfrac12$,
$$
\begin{aligned}
\mathcal A_{\gamma_1}\bigl(\mathcal D_p^{\sigma_x}(\rho)\bigr)
&=\mathrm{diag}(\tfrac34,\tfrac14),\\
\mathcal D_p^{\sigma_x}\bigl(\mathcal A_{\gamma_1}(\rho)\bigr)
&=\mathrm{diag}(\tfrac12,\tfrac12),
\end{aligned}
$$
so the two channels do not commute. The same conclusion holds for $\sigma_y$,
which also exchanges the two energy levels up to phases.

\section{CHSH and CJWR boundary equations}
\label{app:chsh}

We derive the boundary equations used in
Proposition~\ref{prop:windows} and Secs.~\ref{sec:chsh-main}
and~\ref{sec:cjwr-main}. Specifically, the CJWR-$3$ boundary is quadratic,
while the two CHSH boundaries are given by a quadratic and a quartic. We then
identify their physical intervals and intersection.

\paragraph*{{CJWR-$3$ boundary.}}
{Expanding $F_3^2-1=2(Q^2+P^2)+t_z^2-1$ from
Eq.~\eqref{eq:cjwr-F} gives}
\begin{equation}
\kappa_3\lambda^2+2h_3\lambda+\zeta_3=0,
\label{eq:cjwr-complex-quadratic}
\end{equation}
where
\begin{align}
\kappa_3&=2|1-2u|^2+8v^2+(1-\gamma)^2,\nonumber\\
h_3&=4\operatorname{Re}u-8(|u|^2+v^2)+\gamma(1-\gamma),\nonumber\\
\zeta_3&=8(|u|^2+v^2)+\gamma^2-1.
\label{eq:cjwr-complex-coefficients}
\end{align}
Thus $\kappa_3\geq0$.
For a physical $X$ state, equality holds only for
$\sigma_X=\PhiP$. In that case,
$F_3^2\equiv3$, and the CJWR-$3$-satisfying interval is empty.

{When $\kappa_3>0$ and $h_3^2-\kappa_3\zeta_3\geq0$, the two real roots are}
\begin{equation}
R_3^\pm=
\frac{-h_3\pm\sqrt{h_3^2-\kappa_3\zeta_3}}{\kappa_3}.
\label{eq:cjwr-root}
\end{equation}
{The CJWR-$3$-satisfying interval is}
\begin{equation}
\bigl[R_3^-,R_3^+\bigr]\cap[0,1].
\label{eq:cjwr-window}
\end{equation}
{If $h_3^2-\kappa_3\zeta_3<0$, the left-hand side of
Eq.~\eqref{eq:cjwr-complex-quadratic} is positive for every $\lambda$ and the
interval is empty. It is also empty when the roots are real but
$[R_3^-,R_3^+]$ does not meet $[0,1]$.}

\paragraph*{CHSH boundary.}
The CHSH-local interval is $I_{\rm CHSH}=I_2\cap I_4$, where
\begin{align*}
I_2&=\{\lambda\in[0,1]:2(Q^2+P^2)\leq1\},\\
I_4&=\{\lambda\in[0,1]:(Q+P)^2+t_z^2\leq1\}.
\end{align*}
Both defining functions are convex on $[0,1]$, so $I_2$ and $I_4$ are closed
intervals when nonempty.
The boundary equation for $I_2$ is
\begin{equation}
\kappa_Q\lambda^2+2h_Q\lambda+\zeta_Q=0,
\label{eq:chsh-complex-quadratic}
\end{equation}
where
\begin{align}
\kappa_Q&=2|1-2u|^2+8v^2,\nonumber\\
h_Q&=4\operatorname{Re}u-8(|u|^2+v^2),\nonumber\\
\zeta_Q&=8(|u|^2+v^2)-1.
\label{eq:chsh-complex-coefficients}
\end{align}
Here $\kappa_Q\geq0$. If $\kappa_Q=0$, then $I_2$ is empty. When
$\kappa_Q>0$ and
$h_Q^2-\kappa_Q\zeta_Q\geq0$, the two real roots are
\begin{equation}
\widehat q_\pm=
\frac{-h_Q\pm\sqrt{h_Q^2-\kappa_Q\zeta_Q}}{\kappa_Q},
\label{eq:chsh-complex-qroots}
\end{equation}
so $I_2=[\widehat q_-,\widehat q_+]\cap[0,1]$.
If $h_Q^2-\kappa_Q\zeta_Q<0$, the left-hand side of
Eq.~\eqref{eq:chsh-complex-quadratic} is positive for every $\lambda$ and
$I_2$ is empty.

The boundary equation for $I_4$ can be written as
\begin{equation}
2QP=1-Q^2-P^2-t_z^2.
\label{eq:chsh-unsquared}
\end{equation}
Squaring Eq.~\eqref{eq:chsh-unsquared} gives the quartic equation
\begin{equation}
\bigl(Q^2+P^2+t_z^2-1\bigr)^2=4Q^2P^2.
\label{eq:chsh-quartic}
\end{equation}
The quartic has at most four real roots. Only those in $[0,1]$ satisfying
Eq.~\eqref{eq:chsh-unsquared} are retained. Convexity of
$(Q+P)^2+t_z^2$ on $[0,1]$ leaves at most two retained boundary roots. If
there are two, write them as $r_-\leq r_+$ and
$I_4=[r_-,r_+]$. If only $r_+$ is retained, then
$I_4=[0,r_+]$. If no root is retained, $I_4$ is empty.

For real coherences, set $(s_x,s_y,s_z)=(\alpha,\beta,\gamma)$. Then
$t_i(\lambda)=s_i+\lambda(1-s_i)$ for $i\in\{x,y,z\}$. Here $M$ is the
largest of the three pair sums $t_i^2+t_j^2$, so $M\le1$ holds exactly when
all three do. For each $ij\in\{xz,xy,yz\}$, the equation
$t_i^2+t_j^2=1$ takes the form
$\kappa_{ij}\lambda^2+2h_{ij}\lambda+\zeta_{ij}=0$ with
\begin{align*}
\kappa_{ij}&=(1-s_i)^2+(1-s_j)^2,\\
h_{ij}&=s_i(1-s_i)+s_j(1-s_j),\\
\zeta_{ij}&=s_i^2+s_j^2-1 .
\end{align*}
If $\kappa_{ij}=0$ or $h_{ij}^2-\kappa_{ij}\zeta_{ij}<0$ for any pair, the
CHSH-local interval is empty. Otherwise, the real roots are
\begin{equation}
R_{ij}^{\pm}=
\frac{-h_{ij}\pm\sqrt{h_{ij}^2-\kappa_{ij}\zeta_{ij}}}{\kappa_{ij}}.
\label{eq:chsh-real-roots}
\end{equation}
The CHSH-local interval is
\begin{equation}
\bigl[\max_{ij}R^-_{ij},\ \min_{ij}R^+_{ij}\bigr]\cap[0,1].
\label{eq:chsh-window}
\end{equation}
The larger root $R_{ij}^{+}$ is the $R_{ij}$ of
Eq.~\eqref{eq:chsh-min-roots}.

\section{Finite-setting bounds on steerability thresholds}
\label{app:numerics}

We revisit the local-hidden-state model and define the finite measurement
settings. From this formulation, we present the finite-setting semidefinite
program and record the direction sets, solver settings, and numerical checks
used to obtain the reported upper-bound estimates for the
projective-measurement steerability thresholds. The exact steerability results
in Sec.~\ref{sec:true-steering} are analytic and independent of these
calculations.

\paragraph*{Local-hidden-state models.}
Fix a measurement class $\mathcal M$ on $A$. For a measurement
$x\in\mathcal M$ with outcome $a$, the unnormalized conditional state on $B$
is
$$
\sigma_{a|x}=\mathrm{Tr}_A[(E_{a|x}\otimes I)\rho],
$$
where $E_{a|x}\ge0$ and $\sum_aE_{a|x}=I$. The state is unsteerable from
$A$ to $B$ for $\mathcal M$ if there are one-qubit states
$\rho_\mu$, probabilities $p(\mu)$, and response
functions $p(a|x,\mu)$ such that
the following equality holds for every $x\in\mathcal M$ and every $a$:
\begin{equation}
\sigma_{a|x}=\sum_\mu p(\mu)p(a|x,\mu)\rho_\mu.
\label{eq:lhs-model}
\end{equation}
Steering for $\mathcal M$ means that no such model
exists~\cite{CavalcantiSkrzypczyk2017}.

\paragraph*{Measurement classes and threshold order.}
Three measurement classes appear in the comparisons. The smallest is a
fixed finite list $\mathcal M_N$ of projective measurements. For such a list,
Eq.~\eqref{eq:finite-lhs-program} is an exact test, because the encoding
derived below makes it feasible at $\lambda$ precisely when the assemblage of
Eq.~\eqref{eq:finite-assemblage} admits a model of the form of
Eq.~\eqref{eq:lhs-model}, and its optimum is $\lambda^{(N)}_{\rm det}$. The
next class contains all projective qubit measurements and defines
$\lambda_{\rm steer}^{A\to B}$ in Sec.~\ref{sec:true-steering}. The largest
class contains all POVMs. Because the unsteerable sets shrink when the
measurement class grows, the definitive thresholds obey
$$
\lambda_{\rm steer,POVM}^{A\to B}
\le\lambda_{\rm steer}^{A\to B}
\le\lambda^{(N)}_{\rm det}.
$$

The optimized CJWR quantities are witnesses rather than complete tests. A
violation certifies steering, so
$\lambda_{\rm steer}^{A\to B}\le\lambda^{(n)}_{\rm CJWR}$, but nonviolation
does not produce a local-hidden-state model. The CJWR settings are optimized
for each state, whereas $\mathcal M_N$ is held fixed, so no measurement-class
inclusion relates $\lambda^{(N)}_{\rm det}$ to $\lambda^{(n)}_{\rm CJWR}$ and
the numerical order of the two upper bounds can vary along a family.

\paragraph*{Directionality.}
The definitions above treat steering from $A$ to $B$. Exchanging the parties
gives a different local-hidden-state problem, and two-qubit steering can be
one-way~\cite{BowlesVertesiQuintinoBrunner2014}. The optimized CJWR values,
which depend only on the singular values of the correlation tensor, are
symmetric under exchanging the two parties. These witnesses therefore do not
distinguish the two steering directions.

\paragraph*{Finite-label encoding of the finite-setting program.}
We rederive the standard deterministic-table construction that turns
local-hidden-state feasibility into a semidefinite
program~\cite{SkrzypczykNavascuesCavalcanti2014,
CavalcantiSkrzypczyk2017}, so that the description is self-contained.
For the fixed direction set $\mathcal M_N$ of
Sec.~\ref{sec:true-steering}, the measurements are binary and projective, so
the settings are labeled $k=1,\ldots,N$ and the outcomes are
$a\in\{+1,-1\}$. The $2^N$ deterministic outcome tables are the vertices of
the classical response-function polytope for this finite measurement
scenario. We show that every LHS model for the resulting finite assemblage can
be rewritten using these tables and at most $2^N$ hidden states. This is a
finite-setting representation, not a characterization of LHS models for all
projective measurements. The construction encodes Eq.~\eqref{eq:lhs-model} in
the variables of Eq.~\eqref{eq:finite-lhs-program} in two steps.

First, the hidden variable is reduced to a finite label. For the
direction set $\mathcal M_N$, the model of Eq.~\eqref{eq:lhs-model} consists
of probabilities $p(\mu)$, normalized states $\rho_\mu$, and response
functions $p(a|k,\mu)$. Label the deterministic response tables by the strings
$\xi=(\xi_1,\ldots,\xi_N)\in\{+1,-1\}^N$, where $\xi_k$ is the outcome
assigned to setting $k$, and set
$$
q(\xi|\mu)=\prod_{k=1}^{N}p(\xi_k|k,\mu),
\qquad
\sum_{\xi:\,\xi_k=a}q(\xi|\mu)=p(a|k,\mu).
$$
For each $\mu$, the definition on the left is a probability distribution over
the $2^N$ tables, because $\sum_\xi q(\xi|\mu)=1$. The identity on the right
follows by summing out the components except the $k$th, and it shows that
every response function of $\mu$ is a marginal of that distribution.
{The product above is one convenient joint distribution with the required
single-setting marginals. Define}
$$
{p_\xi=\sum_\mu p(\mu)q(\xi|\mu).}
$$
{For $p_\xi>0$, define}
$$
{\tau_\xi=\frac{1}{p_\xi}
\sum_\mu p(\mu)q(\xi|\mu)\rho_\mu.}
$$
{Tables with $p_\xi=0$ are omitted, and they contribute nothing to the
sums below, since $p_\xi=0$ forces $p(\mu)q(\xi|\mu)=0$ for every $\mu$. The
original hidden variable $\mu$ is refined to the pair $(\mu,\xi)$, whose
joint probability is $p(\mu)q(\xi|\mu)$. Its response to setting $k$ is the
predetermined outcome $\xi_k$. For each fixed table $\xi$, summing over the
original hidden labels $\mu$ gives its probability $p_\xi$, and the
corresponding normalized mixture of states $\rho_\mu$ is $\tau_\xi$.
The finite hidden-variable model therefore
reproduces the original LHS assemblage:}
$$
\begin{aligned}
\sum_{\xi:\,\xi_k=a}p_\xi\tau_\xi
&=\sum_\mu p(\mu)\Bigl[\sum_{\xi:\,\xi_k=a}q(\xi|\mu)\Bigr]\rho_\mu\\
&=\sum_\mu p(\mu)p(a|k,\mu)\rho_\mu\\
&{=\sigma_{a|k}.}
\end{aligned}
$$
The first equality uses
$p_\xi\tau_\xi=\sum_\mu p(\mu)q(\xi|\mu)\rho_\mu$, which is the definition of
$\tau_\xi$ with the factor $p_\xi$ cancelled, followed by an exchange of the
two finite sums. The second equality is the marginal identity above, and the
third is Eq.~\eqref{eq:lhs-model} for the original model. The construction
also defines a valid finite hidden-state ensemble $\{p_\xi,\tau_\xi\}$.
Normalization of each $q(\cdot|\mu)$ gives $\sum_\xi p_\xi=1$, and each
$\tau_\xi$ is a normalized positive state because it is a convex mixture of
the states $\rho_\mu$. The shared state $\rho_\lambda$ admits an
$A\to B$ LHS model for $\mathcal M_N$ if and only if there exist probabilities
$p_\xi$ and normalized hidden states $\tau_\xi$ satisfying
$$
\sum_\xi p_\xi\tau_\xi=\rho_B(\lambda),
\qquad
\sum_{\xi:\,\xi_k=+1}p_\xi\tau_\xi=\sigma_{+1|k}(\lambda),
$$
for $k=1,\ldots,N$.

Second, for convenience, combine each probability $p_\xi$ and hidden state
$\tau_\xi$ into the subnormalized hidden state
$\omega_\xi=p_\xi\tau_\xi$. Both conditions above depend on $p_\xi$ and
$\tau_\xi$ only through this product. Treating $\omega_\xi$ as a single
positive semidefinite variable makes the conditions linear and turns the
search into the semidefinite program of Eq.~\eqref{eq:finite-lhs-program}.
The substitution is reversible: every
feasible family $\{\omega_\xi\}$ gives back a model through
$p_\xi=\operatorname{Tr}\omega_\xi$ and $\tau_\xi=\omega_\xi/p_\xi$, with the
zero-trace terms omitted, and the first constraint gives
$\sum_\xi p_\xi=1$.

Only the $+1$ outcome needs to be imposed. Projector completeness gives
$\sigma_{+1|k}(\lambda)+\sigma_{-1|k}(\lambda)=\rho_B(\lambda)$ for
every $k$. Hence, the normalization and positive-outcome constraints imply
$$
\sum_{\xi:\,\xi_k=-1}\omega_\xi
=\rho_B(\lambda)-\sigma_{+1|k}(\lambda)
=\sigma_{-1|k}(\lambda).
$$
Thus, adding the negative-outcome equalities would not change the feasible
set.

Two direction sets are used. They form the nested pair
$\mathcal M_8\subset\mathcal M_{16}$ used for Figs.~\ref{fig:sandwich},
\ref{fig:gap}, and \ref{fig:alignment}. Each listed direction defines one
binary projective measurement. The set $\mathcal M_{16}$ consists
of the directions
$$
\vec n_i=\bigl(\sqrt{1-z_i^2}\cos\phi_i,\,
\sqrt{1-z_i^2}\sin\phi_i,\,z_i\bigr),
$$
where $z_i=(i+\tfrac12)/16$, $\phi_i=\pi(3-\sqrt5)\,i$, and
$i=0,\ldots,15$. The set $\mathcal M_8$ contains the directions with indices
$i\in\{3,4,5,6,7,10,13,14\}$. The opposite direction $-\vec n_i$ is not
listed separately because it defines the same measurement.

The numerical grid contains $39$ noise states for each direction set.
The product-family values are $r_j=j/10$ for $j=0,\ldots,9$. The $r=1$
endpoint shown in Fig.~\ref{fig:sandwich} uses the exact pure-noise result and
is not part of the numerical grid. The two real-coherence grids are
$s_j=j/10$ for $j=0,\ldots,10$.
The phase grid of Sec.~\ref{sec:alignment} is
$\theta_j=j\pi/6$ for $j=0,\ldots,6$, evaluated at the fixed coherence
fraction $s=0.95$ on the $u=v$ family and used for
Fig.~\ref{fig:alignment}. Solving the finite-setting program for both
$\mathcal M_8$ and $\mathcal M_{16}$ gives $78$ estimates in total.

The numerical calculation solves Eq.~\eqref{eq:finite-lhs-program} with
CVXPY 1.9.2~\cite{DiamondBoyd2016CVXPY}. The program carries one hidden operator per response table, so its
size grows as $2^N$.
The programs use primal-dual gap and feasibility tolerances of $10^{-8}$.
All reported values $\widehat{\lambda}^{(N)}_{\rm det}$ were produced by
CLARABEL 0.11.1~\cite{GoulartChen2024Clarabel}. As a solver check, all $78$ programs were repeated with
MOSEK~\cite{MOSEK112}. The
largest absolute difference between the CLARABEL and MOSEK estimates is
$1.61\times10^{-6}$, and every estimate is unchanged through three decimal
places. At the sixteen-direction Werner point, the
values are $0.5110647$ for CLARABEL and $0.5110648$ for MOSEK, a difference of
$8.4\times10^{-8}$.

\paragraph*{Numerical checks for the finite-setting estimates.}
The computed estimates satisfy
$\widehat{\lambda}^{(16)}_{\rm det}\le\widehat{\lambda}^{(8)}_{\rm det}$ at
all $39$ points. For each reported CLARABEL estimate, the returned hidden
operators $\{\omega_\xi\}$ are substituted into the constraints of
Eq.~\eqref{eq:finite-lhs-program} and evaluated in floating-point arithmetic.
Across the $78$ reported CLARABEL solutions, the maximum absolute entrywise
equality-constraint residual is below $7\times10^{-10}$, and every hidden
operator has minimum eigenvalue above $-6\times10^{-10}$. These residuals
show numerical consistency with a local-hidden-state decomposition at the
reported value.

\section{Relative-phase monotonicity from the explicit boundary conditions}
\label{app:mono}

We present the proofs of the inclusions stated in
Sec.~\ref{sec:alignment} for the CHSH-local, CJWR-$n$-satisfying,
teleportation-useless, and separable intervals. Each proof follows directly
from the defining condition of the corresponding interval
and from how that condition depends
on the relative phase
$\theta=\arg u$. Finally, we list the noise parameters used in
Fig.~\ref{fig:gallery}.
Throughout, $x=\cos\theta$. When two phases are compared, we assume
$\pi\ge\theta_1\ge\theta_2\ge0$, so that
$-1\le x_1\le x_2\le1$.

\subsection{{CHSH-local and CJWR-satisfying intervals}}

{With $a,b,c,d,|u|$, and $v$ fixed, Proposition~\ref{prop:flow} gives}
$$
Q(\lambda,x)^2=\lambda^2+4\lambda(1-\lambda)|u|x
+4(1-\lambda)^2|u|^2 .
$$
If $x_1\le x_2$, then
$$
Q(\lambda,x_2)^2-Q(\lambda,x_1)^2
=4\lambda(1-\lambda)|u|(x_2-x_1)\ge0
$$
for every $\lambda\in[0,1]$, and since $Q\ge0$, also
$Q(\lambda,x_1)\le Q(\lambda,x_2)$. {The other two quantities that
determine the correlation-tensor singular values,
$P(\lambda)=2(1-\lambda)v\ge0$ and $t_z(\lambda)$, do not depend on $x$.}
Write the two branches of the optimized CHSH quantity
{in Eq.~\eqref{eq:chsh-max}} as
$$
\begin{aligned}
{H_1(\lambda,x)}&{=2\bigl[Q(\lambda,x)^2+P(\lambda)^2\bigr]},\\
{H_2(\lambda,x)}&{=\bigl[Q(\lambda,x)+P(\lambda)\bigr]^2
+t_z(\lambda)^2}.
\end{aligned}
$$
Because $P\ge0$, the sum $Q+P$ is nonnegative, so both $H_1$ and $H_2$ are
nondecreasing functions of $Q$ on $Q\ge0$. Hence,
$H_j(\lambda,x_1)\le H_j(\lambda,x_2)$ for $j=1,2$, and therefore
$$
{M(\lambda,x_1)
=\max\bigl\{H_1(\lambda,x_1),H_2(\lambda,x_1)\bigr\}
\le M(\lambda,x_2).}
$$
If a Bell weight $\lambda$ is CHSH-local at $x_2$, then
$$
M(\lambda,x_1)\le M(\lambda,x_2)\le1,
$$
so it is CHSH-local at $x_1$. Hence, the CHSH-local set at $x_2$ is contained
in the CHSH-local set at $x_1$.
By Proposition~\ref{prop:windows}, each of these sets is a closed
interval, possibly empty. {Whenever the interval at $x_2$ is nonempty,
both intervals are nonempty, and the inclusion implies that the lower edge
cannot fall and the upper edge cannot rise when $x$ increases.}
Equivalently, neither edge of the CHSH-local interval moves inward as
$\theta$ runs from $0$ to $\pi$. Eq.~\eqref{eq:cjwr-F}, together
with $F_n\ge0$, likewise shows that $F_n(\lambda,x)$ is nondecreasing in $x$
for $n\in\{2,3\}$. Thus, if $\lambda$ is CJWR-$n$-satisfying at $x_2$, then
$$
F_n(\lambda,x_1)\le F_n(\lambda,x_2)\le1,
$$
so it is CJWR-$n$-satisfying at $x_1$ as well. With
Proposition~\ref{prop:windows} applied as above, this proves that the
CJWR-$n$-satisfying interval widens as $\theta$ increases.

\subsection{Teleportation usefulness}

In the variable $\eta=\lambda/(1-\lambda)$, the phase-dependent condition
in Eq.~\eqref{eq:eta-teleportation} is
$$
\eta^2+4\eta|u|x+4|u|^2\le (b+c)^2.
$$
For $x_1\le x_2$ and every $\eta\ge0$, if this condition holds at $x_2$,
then it also holds at $x_1$, since the left-hand side is nondecreasing in
$x$:
$$
\eta^2+4\eta|u|x_1+4|u|^2
\le\eta^2+4\eta|u|x_2+4|u|^2\le (b+c)^2.
$$
The other condition, $\eta\ge2|v|-(a+d)$, does not depend on $x$.
Therefore, at fixed $\eta$, teleportation uselessness at $x_2$ implies
teleportation uselessness at $x_1$. Since $\lambda=\eta/(1+\eta)$ is strictly
increasing in $\eta$, the teleportation-useless interval widens as $\theta$
increases.

\subsection{Separability}

For the separable interval, the only phase-dependent PPT
condition in Eq.~\eqref{eq:eta-ppt} is
$$
|u|^2+\eta|u|x+\frac{\eta^2}{4}\le bc.
$$
Its left-hand side is again nondecreasing in $x$, so the argument just
given for the teleportation-useless interval shows that the separable interval
widens as $\theta$ increases.

Each of these inclusions transfers to the corresponding definitive
threshold of Eq.~\eqref{eq:definitive}. When the interval at $x_2$ is
nonempty, it is contained in the interval at $x_1$, so its supremum is no
larger. When the interval at $x_2$ is empty, its definitive threshold is zero
and the ordering holds trivially. The five definitive thresholds of
Theorem~\ref{thm:mono} are therefore nonincreasing in $x$, equivalently
nondecreasing in $\theta$ on $[0,\pi]$.

\subsection{Noise states used in the operational phase diagrams}

Each of the six panels~(a)--(f) of Fig.~\ref{fig:gallery} varies $\theta$
in $u=|u|e^{i\theta}$ while holding $a$, $b$, $c$, $d$, $|u|$, and $v\ge0$ at
the values below:
$$
\begin{array}{c|cccccc}
\text{panel} & a & b & c & d & |u| & v\\
\hline
(a) & 0.405 & 0 & 0.190 & 0.405 & 0.395 & 0\\
(b) & 0.400  & 0      & 0.200 & 0.400  & 0.260 & 0\\
(c) & 0.020  & 0.120  & 0.560 & 0.300  & 0.060 & 0.120\\
(d) & 0.340  & 0.080  & 0.578 & 0.002  & 0.014 & 0.210\\
(e) & 0.230 & 0.520 & 0.230 & 0.020 & 0.065 & 0.330\\
(f) & 0.210 & 0.490 & 0.250 & 0.050 & 0.100 & 0.345
\end{array}
$$
{For every row, $a+b+c+d=1$, $|u|^2\le ad$, and $v^2\le bc$, so the
listed parameters define a physical $X$ state. Rows~(a) and~(b) satisfy
$|u|^2>bc$ and therefore have entanglement in the $\Phi$ block. Rows~(c)--(f)
satisfy $v^2>ad$ and therefore have entanglement in the $\Psi$ block.}

\section{Product noise}
\label{app:product}

We first prove Proposition~\ref{prop:product} by deriving the
singular values of the correlation tensor of the Bell mixture for the case of
arbitrary product noise. From these singular values, we obtain the
CJWR-$3$ witness threshold, the fully entangled fraction, and the teleportation
usefulness threshold. The last formula reproduces the arbitrary-product
result of Ref.~\cite{eac-paper}. We also record the CHSH and CJWR witness
thresholds on the aligned product family. We close by proving
Proposition~\ref{prop:product-pure-factor}, which gives the exact steerability
thresholds for product noise with at least one pure local factor.

\paragraph*{Singular-value flow.}
Let $\sigma=\sigma_A\otimes\sigma_B$ have Bloch vectors
$\vec r_A=r_A\vec n_A$ and $\vec r_B=r_B\vec n_B$. As in
Proposition~\ref{prop:product}, $D=\mathrm{diag}(1,-1,1)$ is the correlation
tensor of $\PhiP$. Write $T_\lambda=T(\rho_\lambda)$. The
correlation tensor of $\rho_\lambda$ is
$$
T_\lambda=\lambda D+(1-\lambda)r_Ar_B\,
\vec n_A\vec n_B^{\mathsf T}.
$$
Consequently, $T_\lambda T_\lambda^{\mathsf T}=\lambda^2I+K$, where
$K$ is symmetric, has rank at most two, and equals
$$
K=\mathcal J(\vec m_B\vec n_A^{\mathsf T}
+\vec n_A\vec m_B^{\mathsf T})
+\mathcal L\,\vec n_A\vec n_A^{\mathsf T}.
$$
Here, $\vec m_B=D\vec n_B$, $\mathcal J=\lambda(1-\lambda)g$,
$\mathcal L=(1-\lambda)^2g^2$, and $g=r_Ar_B$.
Proposition~\ref{prop:product} defines
$\cos\varphi=\vec n_A\cdot\vec m_B$.
\mbox{If $\sin\varphi\ne0$, set}\nopagebreak[4]
$$
\vec e_1=\vec n_A,
\qquad
\vec e_2=\frac{\vec m_B-\cos\varphi\,\vec n_A}{\sin\varphi}.
$$
When $\sin\varphi=0$, choose any unit vector $\vec e_2$ orthogonal to
$\vec e_1$. On the plane spanned by $\vec e_1$ and $\vec e_2$,
$$
\left.K\right|_{\mathrm{span}\{\vec e_1,\vec e_2\}}=
\begin{pmatrix}
2\mathcal J\cos\varphi+\mathcal L & \mathcal J\sin\varphi\\
\mathcal J\sin\varphi & 0
\end{pmatrix}.
$$
Its trace is $2\mathcal J\cos\varphi+\mathcal L$ and its determinant is
$-\mathcal J^2\sin^2\varphi$, which are, respectively, the sum and product of
the two eigenvalues $\kappa_\pm$ of this restriction. Since $K$ vanishes on
the orthogonal complement of this plane, its third eigenvalue is zero. Thus,
$$
\kappa_\pm=
\frac{2\mathcal J\cos\varphi+\mathcal L
\pm\sqrt{(2\mathcal J\cos\varphi+\mathcal L)^2
+4\mathcal J^2\sin^2\varphi}}{2}.
$$
Since $T_\lambda T_\lambda^{\mathsf T}=\lambda^2I+K$, the singular
values of $T_\lambda$ are $\lambda$ and
$s_\pm=\sqrt{\lambda^2+\kappa_\pm}$.
Using $\kappa_++\kappa_-=2\mathcal J\cos\varphi+\mathcal L$,
$\kappa_+\kappa_-=-\mathcal J^2\sin^2\varphi$, and
$\mathcal J^2=\lambda^2\mathcal L$, we obtain
$$
\begin{aligned}
s_+^2s_-^2
&=(\lambda^2+\kappa_+)(\lambda^2+\kappa_-)\\
&=\lambda^4+\lambda^2(2\mathcal J\cos\varphi+\mathcal L)
-\mathcal J^2\sin^2\varphi\\
&=(\lambda^2+\mathcal J\cos\varphi)^2.
\end{aligned}
$$
Because $s_\pm\ge0$ and $\lambda\ge0$, the preceding identity and the
sum $\kappa_++\kappa_-$ give
$$
\begin{aligned}
s_+s_-
&=|\lambda^2+\mathcal J\cos\varphi|
=\lambda|\lambda+(1-\lambda)g\cos\varphi|,\\
s_+^2+s_-^2
&=2\lambda^2+\kappa_++\kappa_-
=2\lambda^2+2\mathcal J\cos\varphi+\mathcal L.
\end{aligned}
$$
Consequently,
$$
(s_+\pm s_-)^2
=2\lambda^2+2\mathcal J\cos\varphi+\mathcal L
\pm2|\lambda^2+\mathcal J\cos\varphi|.
$$
$$
\bigl\{(s_++s_-)^2,(s_+-s_-)^2\bigr\}
=\bigl\{4\lambda^2+4\mathcal J\cos\varphi+\mathcal L,
\mathcal L\bigr\}.
$$
Finally, substituting
$\mathcal J=\lambda(1-\lambda)g$ and
$\mathcal L=(1-\lambda)^2g^2$ gives
$$
\begin{aligned}
4\lambda^2+4\mathcal J\cos\varphi+\mathcal L
&=|2\lambda+(1-\lambda)ge^{i\varphi}|^2,\\
\mathcal L&=(1-\lambda)^2g^2.
\end{aligned}
$$
Define
$\widetilde Q=|\lambda+(1-\lambda)(g/2)e^{i\varphi}|$ and
$\widetilde P=(1-\lambda)g/2$. The two singular values are then
$$
s_\pm=|\widetilde Q\pm\widetilde P|.
$$
Together with the remaining singular value $\lambda$, these expressions
agree with the effective $X$ flow of Eq.~\eqref{eq:sv-flow} for
$u_{\rm eff}=(g/4)e^{i\varphi}$ and $v_{\rm eff}=g/4$. The third singular
value equals $\lambda$ when $\gamma_{\rm eff}=0$, which the populations
$a_{\rm eff}=b_{\rm eff}=c_{\rm eff}=d_{\rm eff}=1/4$ realize. Since $g\le1$,
they also satisfy $|u_{\rm eff}|^2\le a_{\rm eff}d_{\rm eff}$ and
$v_{\rm eff}^2\le b_{\rm eff}c_{\rm eff}$, so the effective state is a
physical $X$ state and Theorem~\ref{thm:mono} applies.
At fixed $g$,
Appendix~\ref{app:mono} then shows that the CHSH-nonlocal, CJWR witness, and
teleportation usefulness thresholds are nonincreasing in $\cos\varphi$.

\paragraph*{CJWR-$3$ witness threshold.}
Using the three singular values above, the three-setting CJWR expression
is
\begin{equation}
F_3^2(\lambda)
=3\lambda^2+2g\cos\varphi\,\lambda(1-\lambda)
+g^2(1-\lambda)^2.
\label{eq:product-cjwr3-value}
\end{equation}
At $\lambda=0$, $\rho_0$ is a product state and therefore unsteerable,
so it satisfies the CJWR-$3$ inequality. This also follows from
$F_3^2(0)-1=g^2-1\le0$. The boundary equation
$F_3^2(\lambda)=1$ is
$$
(3-2g\cos\varphi+g^2)\lambda^2
+2g(\cos\varphi-g)\lambda+g^2-1=0.
$$
Its leading coefficient is
$3-2g\cos\varphi+g^2\ge(g-1)^2+2>0$. Thus, the product of the roots is
nonpositive, so at most one root is positive. Since $F_3^2(1)-1=2$, the
larger root lies in $[0,1]$ and is the CJWR-$3$ witness threshold.
The discriminant is $4\mathcal V_3(g,\varphi)$, where
\begin{equation}
\mathcal V_3(g,\varphi)
=3-2g\cos\varphi-2g^2+g^2\cos^2\varphi.
\label{eq:product-cjwr3-discriminant}
\end{equation}
The larger root is
\begin{equation}
\lambda^{(3)}_{\rm CJWR}(g,\varphi)
=\frac{
g(g-\cos\varphi)
+\sqrt{\mathcal V_3(g,\varphi)}
}{3-2g\cos\varphi+g^2}.
\label{eq:product-cjwr3-threshold}
\end{equation}

\paragraph*{Fully entangled fraction and teleportation usefulness threshold.}
Let $s_{(1)}\ge s_{(2)}\ge s_{(3)}$ be the ordered singular values of
the correlation tensor $T$ of a two-qubit state. The fully entangled fraction
can be written as~\cite{HorodeckiTeleportationBell1996,
HorodeckiTeleportation1999}
$$
f=\frac14\left[1+s_{(1)}+s_{(2)}
-\operatorname{sgn}(\det T)s_{(3)}\right].
$$
For the product-noise mixture, left multiplication by
$D$ and the matrix determinant lemma give
$$
\det T_\lambda
=-\lambda^2\bigl[\lambda+(1-\lambda)g\cos\varphi\bigr].
$$
Moreover,
$$
\widetilde Q^2-\widetilde P^2
=\lambda\bigl[\lambda+(1-\lambda)g\cos\varphi\bigr].
$$
The two displays give
$\det T_\lambda=-\lambda(\widetilde Q^2-\widetilde P^2)$. If
$\widetilde Q\ge\widetilde P$, the determinant is nonpositive, so the bracket
in $f$ adds all three singular values and equals
$\lambda+2\widetilde Q$. If
$\widetilde P>\widetilde Q$, the determinant is positive, and subtracting
the smaller of $\lambda$ and $\widetilde P-\widetilde Q$ gives the larger of
$\lambda+2\widetilde Q$ and $2\widetilde P-\lambda$. Thus,
\begin{equation}
f(\rho_\lambda)
=\frac14\left[1+\max\left\{
\lambda+2\widetilde Q,\;2\widetilde P-\lambda
\right\}\right].
\label{eq:product-fully-entangled-fraction}
\end{equation}

Because $g\le1$ gives $2\widetilde P\le1-\lambda$, the second entry obeys
$2\widetilde P-\lambda\le1$. Hence,
Eq.~\eqref{eq:product-fully-entangled-fraction} reduces
$f(\rho_\lambda)>1/2$ to $\lambda+2\widetilde Q>1$.
With $\eta=\lambda/(1-\lambda)$, the boundary equality
$\lambda+2\widetilde Q=1$ becomes
$2|\eta+(g/2)e^{i\varphi}|=1$. Squaring gives
$$
4\eta^2+4g\cos\varphi\,\eta+g^2=1.
$$
Its larger root is
$$
\eta_F
=\frac{\sqrt{1-g^2\sin^2\varphi}-g\cos\varphi}{2}.
$$
Since $\lambda=0$ is separable and therefore teleportation-useless,
$\eta_F$ is the upper edge of the teleportation-useless interval. The
strictly increasing relation $\lambda=\eta/(1+\eta)$ then gives
\begin{equation}
\lambda_F(g,\varphi)
=\frac{\sqrt{1-g^2\sin^2\varphi}-g\cos\varphi}
{2+\sqrt{1-g^2\sin^2\varphi}-g\cos\varphi}.
\label{eq:product-teleportation-threshold}
\end{equation}
The teleportation usefulness threshold in
Eq.~\eqref{eq:product-teleportation-threshold} agrees with the
arbitrary-product result derived in Ref.~\cite{eac-paper}.

\paragraph*{CHSH and CJWR witness thresholds on aligned products.}
For $\sigma_A=\sigma_B=(I+r\sigma_z)/2$, the Bloch vectors are
$\vec r_A=\vec r_B=r\vec e_z$, so $g=r^2$ and $\cos\varphi=1$, and the
correlation tensor is
$$
T_\lambda
=\mathrm{diag}(\lambda,-\lambda,\lambda+(1-\lambda)r^2).
$$
The two closed-form thresholds plotted in Fig.~\ref{fig:sandwich} are
$$
\lambda_{\mathrm{CHSH}}(r)
=\frac{\sqrt{2(1-r^2)}-r^2(1-r^2)}
{1+(1-r^2)^2}
$$
and
$$
\lambda^{(3)}_{\rm CJWR}(r)
=\frac{\sqrt{(1-r^2)(3+r^2)}-r^2(1-r^2)}
{2+(1-r^2)^2}.
$$
The second formula is Eq.~\eqref{eq:product-cjwr3-threshold} at
$g=r^2$ and $\cos\varphi=1$, while the first is the root in $[0,1]$ of
$M(\lambda)=1$ for the displayed tensor.
At $r=0$, the formulas recover
$\lambda_{\rm CHSH}=1/\sqrt2$, $\lambda^{(3)}_{\rm CJWR}=1/\sqrt3$, and
$\lambda_*=\lambda_F=1/3$. At $r=1$, all four thresholds vanish.
Proposition~\ref{prop:product} extends both formulas to aligned product
noise with unequal Bloch radii by replacing $r^2$ with $g=r_Ar_B$.
Thus, every threshold determined by the correlation tensor, including
$\lambda_F$ of Eq.~\eqref{eq:product-teleportation-threshold}, depends only
on $g$, whereas the
entanglement threshold
$$
\lambda_*=\frac{\sqrt{(1-r_A^2)(1-r_B^2)}}
{2+\sqrt{(1-r_A^2)(1-r_B^2)}}
$$
depends on $r_A$ and $r_B$ separately~\cite{eac-paper}.

\begingroup

\paragraph*{Proof of Proposition~\ref{prop:product-pure-factor}.}
Assume first that $\sigma_A=|a\rangle\!\langle a|$. Choose a unitary
$U$ such that $U|a\rangle=|0\rangle$. Since
$(U\otimes U^*)|\Phi^+\rangle=|\Phi^+\rangle$, the local unitary
$U\otimes U^*$ leaves the Bell term unchanged. Concurrence and
steerability in either direction are invariant under local unitaries, so we
may take $\sigma_A=|0\rangle\!\langle0|$
and write the transformed state of $B$ as
$$
\sigma_B=
\begin{pmatrix}
p&w\\
w^*&1-p
\end{pmatrix}.
$$
Here, $0\le p\le1$ and $|w|^2\le p(1-p)$.
In the computational basis, the Bell mixture is then
$$
\rho_\lambda=
\begin{pmatrix}
\lambda/2+(1-\lambda)p&(1-\lambda)w&0&\lambda/2\\
(1-\lambda)w^*&(1-\lambda)(1-p)&0&0\\
0&0&0&0\\
\lambda/2&0&0&\lambda/2
\end{pmatrix}.
$$
Define the spin flip by
$\widetilde\rho_\lambda=(\sigma_y\otimes\sigma_y)
\rho_\lambda^*(\sigma_y\otimes\sigma_y)$.
Although $\rho_\lambda\widetilde\rho_\lambda$ need not be Hermitian,
it has the same characteristic polynomial as
$\sqrt{\rho_\lambda}\,\widetilde\rho_\lambda\sqrt{\rho_\lambda}$,
which is positive semidefinite. Consequently, the eigenvalues of
$\rho_\lambda\widetilde\rho_\lambda$ are real and nonnegative. Let
$\xi_1\ge\xi_2\ge\xi_3\ge\xi_4\ge0$ be the eigenvalues of
$\rho_\lambda\widetilde\rho_\lambda$. Wootters'
formula~\cite{Wootters1998} gives
$$
C(\rho_\lambda)
=\max\{0,\sqrt{\xi_1}-\sqrt{\xi_2}-\sqrt{\xi_3}-\sqrt{\xi_4}\}.
$$
Writing $\bar\lambda=1-\lambda$, direct multiplication gives
$$
\rho_\lambda\widetilde\rho_\lambda
=\frac{\lambda}{2}
\begin{pmatrix}
\lambda+\bar\lambda p&0&-\bar\lambda w^*&\lambda+2\bar\lambda p\\
\bar\lambda w^*&0&0&\bar\lambda w^*\\
0&0&0&0\\
\lambda&0&-\bar\lambda w^*&\lambda+\bar\lambda p
\end{pmatrix}.
$$
Define the characteristic polynomial by
$\mathcal P(\xi)=\det(\xi I-\rho_\lambda\widetilde\rho_\lambda)$. Its
explicit form is
$$
\mathcal P(\xi)=\xi^2\!\left[\xi^2
-\bigl(\lambda^2+\lambda(1-\lambda)p\bigr)\xi
+\tfrac14\lambda^2(1-\lambda)^2p^2\right].
$$
The roots of $\mathcal P$ are $\xi_i$. Thus,
$\xi_3=\xi_4=0$, while the sum and product of the two remaining roots give
$$
\xi_1+\xi_2=\lambda^2+\lambda(1-\lambda)p,
\qquad
\xi_1\xi_2=\frac{\lambda^2(1-\lambda)^2p^2}{4}.
$$
It follows that
$(\sqrt{\xi_1}-\sqrt{\xi_2})^2=\lambda^2$. Since
$\xi_1\ge\xi_2$ and $\lambda\ge0$, Wootters' formula yields
$$
C(\rho_\lambda)=\sqrt{\xi_1}-\sqrt{\xi_2}=\lambda
$$
for every $\lambda\in[0,1]$. Thus, $\rho_\lambda$ is entangled exactly
when $\lambda>0$, and $\lambda_*=0$. It remains to
evaluate the steerability thresholds. If $\sigma_B$ has rank one, then
$\sigma_A\otimes\sigma_B$ is pure, and
Theorem~\ref{cor:pure-noise-steering} gives all four steerability thresholds
equal to $\lambda_*=0$.

Suppose instead that $\sigma_B$ has rank two. For every $\lambda>0$,
the matrix above gives
$$
\rho_\lambda|1,0\rangle=0.
$$
In Lemma~\ref{lem:product-null}, take
$|e,f\rangle=|1,0\rangle$, so that
$|e^\perp,f\rangle=|0,0\rangle$ and
$|e,f^\perp\rangle=|1,1\rangle$. The required matrix element is
$$
\langle e^\perp,f|\rho_\lambda|e,f^\perp\rangle
=\langle0,0|\rho_\lambda|1,1\rangle
=\frac{\lambda}{2}\ne0.
$$
Lemma~\ref{lem:product-null} therefore gives steerability in both
directions using projective measurements for every $\lambda>0$.
The same proof applies when $\sigma_B$ is the pure factor.

At $\lambda=0$, $\rho_0=\sigma_A\otimes\sigma_B$ is separable and
unsteerable for every measurement class. Since projective measurements are
contained among arbitrary POVMs, projective-measurement steerability for
every $\lambda>0$ also gives POVM steerability for every $\lambda>0$.
Therefore, $\lambda=0$ is the unique separable Bell weight and, in either
direction, the unique Bell weight that is unsteerable under projective
measurements or arbitrary POVMs.
This proves Eq.~\eqref{eq:product-pure-factor-thresholds}.
\hfill$\square$
\endgroup

\section{Post-selected swapping chain}
\label{app:swapping}

We derive the recurrence for the swapping chain of
Sec.~\ref{sec:death-times} and show why teleportation usefulness and
entanglement are lost at the same swap depth for that sequence.

The trajectory in Fig.~\ref{fig:death}(b) starts from
$\rho_0$ in Eq.~\eqref{eq:swap-link}. At each step, the
state $\rho_n^{12}$ is combined with an elementary link $\rho_0^{34}$, and
the Bell-measurement outcome $\PhiP_{23}$ is retained. The
unnormalized state of the two outer qubits is
$$
\rho'_{n+1}
={}_{23}\!\langle\Phi^+|
\bigl(\rho_n^{12}\otimes\rho_0^{34}\bigr)
|\Phi^+\rangle_{23}.
$$
The probability of this outcome is
$p_{n+1}=\operatorname{Tr}\rho'_{n+1}$, and
$\rho_{n+1}=\rho'_{n+1}/p_{n+1}$. Thus, after $n$ swaps, the
process has used $n+1$ elementary links and $n$ Bell measurements.

For the elementary link obtained by local amplitude damping in
Eq.~\eqref{eq:swap-link}, with $\gamma_1=0.95$,
$$
\begin{aligned}
a_0&=\tfrac12\bigl[1+(1-\gamma_1)^2\bigr],\\
b_0=c_0&=\tfrac12\gamma_1(1-\gamma_1),
&d_0&=\tfrac12\gamma_1^2,\\
u_0&=\tfrac12\gamma_1,
&v_0&=0.
\end{aligned}
$$
Write the entries of $\rho_n$ as
$(a_n,b_n,c_n,d_n,u_n,v_n)$, following Eq.~\eqref{eq:x-matrix}. Direct
evaluation of the post-selected state gives
$$
\begin{aligned}
a'_{n+1}&=\tfrac12(a_na_0+b_nc_0),
&b'_{n+1}&=\tfrac12(a_nb_0+b_nd_0),\\
c'_{n+1}&=\tfrac12(c_na_0+d_nc_0),
&d'_{n+1}&=\tfrac12(c_nb_0+d_nd_0),\\
u'_{n+1}&=\tfrac12u_nu_0,
&v'_{n+1}&=\tfrac12v_nu_0.
\end{aligned}
$$
Here,
$p_{n+1}=a'_{n+1}+b'_{n+1}
+c'_{n+1}+d'_{n+1}$, and the normalized entries are
obtained by dividing the six quantities above by $p_{n+1}$. The initial
entries satisfy $b_0=c_0$, $v_0=0$, and
$a_0-d_0=2b_0/\gamma_1$. The recurrence therefore shows by induction that
$$
b_n=c_n,\qquad v_n=0,\qquad u_n\ge0,
\qquad a_n-d_n=\frac{2b_n}{\gamma_1}
$$
for every $n$. Thus, this particular sequence remains in the symmetric
real-$X$ family. For these outputs, the $X$-state concurrence formula and
the fully entangled-fraction formula give
$$
\begin{aligned}
C(\rho_n)&=2\max\{0,u_n-b_n\},\\
f(\rho_n)&=\max\!\left\{\tfrac12+u_n-b_n,b_n\right\}.
\end{aligned}
$$
Because $b_n\le\tfrac12$, the second branch of $f(\rho_n)$ never exceeds
$1/2$. Therefore,
$$
C(\rho_n)=2\max\{0,f(\rho_n)-\tfrac12\}
$$
for every $n$. Standard teleportation usefulness and entanglement are
therefore lost together along this sequence.

Iterating the recurrence and evaluating the four functions in
Eq.~\eqref{eq:loss-functions} after every swap gives the ability-loss depths
$n_{\rm CHSH}=6$, $n_{\mathrm{CJWR}\text{-}3}=8$, and $n_F=n_*=16$, as
quoted in Sec.~\ref{sec:death-times}. Figure~\ref{fig:death}(b) displays these
functions for the post-selected state $\rho_n$. It does not include the
cumulative success probability
$P^{(n)}_{\rm succ}=\prod_{k=1}^{n}p_k$ of the $n$ Bell
measurements. Before each Bell measurement, the reduced state of the two
measured qubits is a product state. Therefore, the probability of the
$\PhiP$ outcome satisfies $p_k\le1/2$, and
$P^{(n)}_{\rm succ}\le2^{-n}$.

\bibliographystyle{apsrev4-2}
\bibliography{refs}

\end{document}